\documentclass[12pt]{article}

\usepackage{epsfig}
\usepackage{amssymb}
\usepackage{amsbsy}
\usepackage{amsmath}
\usepackage{amsfonts}
\usepackage{mathrsfs}
\usepackage{mathtools}

\newcommand{\R}{\mathbb{R}}
\newcommand{\C}{\mathbb{C}}

\newcommand{\bfA}{\mathbf{A}}
\newcommand{\bbfA}{{\boldsymbol{\mathcal{A}}}}

\newcommand{\bbe}{{\mathbf{e}}}
\newcommand{\bfE}{\mathbf{E}}
\newcommand{\bfB}{\mathbf{B}}
\newcommand{\bfS}{\mathbf{S}}

\newcommand{\bU}{\mathbf{U}}
\newcommand{\bfV}{\mathbf{V}}

\newcommand{\bfk}{{\mathbf{k}}}
\newcommand{\bfm}{\mathbf{m}}
\newcommand{\bfn}{\mathbf{n}}
\newcommand{\bfu}{\mathbf{u}}
\newcommand{\bfx}{\mathbf{x}}
\newcommand{\bfy}{\mathbf{y}}

\newcommand{\bfp}{\mathbf{p}}

\newcommand{\bfX}{\mathbf{X}}
\newcommand{\bfP}{\mathbf{P}}
\newcommand{\bfQ}{\mathbf{Q}}

\newcommand{\bfa}{\mathbf{a}}

\newcommand{\bfr}{\mathbf{r}}

\newcommand{\bone}{\mathbf{1}}
\newcommand{\bfxi}{{\boldsymbol{\xi}}}

\newcommand{\bfzeta}{{\boldsymbol{\zeta}}}

\newcommand{\fm}{\mathfrak{m}}

\newcommand{\fU}{\mathfrak{U}}

\newcommand{\fh}{{\mathfrak{h}}}
\newcommand{\Lfh}{{\mbox{\large$\mathfrak{h}$}}}

\newcommand{\bK}{\mathbf{K}}

\newcommand{\fX}{\mathfrak{X}}

\newcommand{\be}{\begin{equation}}
\newcommand{\ee}{\end{equation}}
\newcommand{\bea}{\begin{eqnarray}}
\newcommand{\eea}{\end{eqnarray}}

\newcommand{\nn}{\nonumber}
\newcommand{\kt}{\rangle}
\newcommand{\br}{\langle}

\newcommand{\ed}{\end{document}}
\newcommand{\bbr}{\br\!\br}
\newcommand{\kkt}{\kt\!\kt}
\newcommand{\pbr}{\prec\!}
\newcommand{\pkt}{\!\succ}

\newcommand{\cbr}{(\!(}
\newcommand{\ckt}{)\!)}
\newcommand{\f}{\frac}

\newcommand{\lnd}{\boldsymbol{\lambda}}
\newcommand{\ep}{\epsilon}
\newcommand{\cH}{\mathcal{H}}
\newcommand{\cJ}{\mathcal{J}}
\newcommand{\cK}{\mathcal{K}}

\newcommand{\cB}{\mathcal{B}}

\newcommand{\cE}{\mathcal{E}}
\newcommand{\cO}{\mathcal{O}}
\newcommand{\cV}{\mathcal{V}}

\newcommand{\cX}{\mathcal{X}}
\newcommand{\cY}{\mathcal{Y}}
\newcommand{\cU}{\mathcal{U}}
\newcommand{\cS}{\mathcal{S}}

\newcommand{\bfv}{{\mbox{v}}}
\newcommand{\bsigma}{\boldsymbol{\sigma}}

\newcommand{\blambda}{\boldsymbol{\lambda}}
\newcommand{\bnabla}{\boldsymbol{\nabla}}

\newcommand{\sH}{\mathscr{H}}

\newcommand{\RE}{{\rm Re}}

\begin{document}

\title{Quantum Mechanics of a Photon}

\author{Hassan Babaei$^{1}$ and Ali~Mostafazadeh$^{1,2,}$\thanks{Corresponding Author, E-mail address:
amostafazadeh@ku.edu.tr}\\[6pt]
Departments of Physics$^{1}$ and Mathematics$^{2}$, Ko\c{c} University,\\ 34450 Sar{\i}yer,
Istanbul, Turkey}

\date{Resubmission date: July 26, 2017}
\maketitle

\begin{abstract}

A first quantized free photon is a complex massless vector field $A=(A^\mu)$ whose field strength satisfies Maxwell's equations in vacuum. We construct the Hilbert space $\sH$ of the photon by endowing the vector space of the fields $A$ in the temporal-Coulomb gauge with a positive-definite and relativistically invariant inner product. We give an explicit expression for this inner product, identify the Hamiltonian for the photon with the generator of time translations in $\sH$, determine the operators representing the momentum and the helicity of the photon, and introduce a chirality operator whose eigenfunctions correspond to fields having a definite sign of energy. We also construct a position operator for the photon whose components commute with each other and with the chirality and helicity operators. This allows for the construction of the localized states of the photon with a definite sign of energy and helicity. We derive an explicit formula for the latter and compute the corresponding electric and magnetic fields. These turn out to diverge not just at the point where the photon is localized but on a plane containing this point. We identify the axis normal to this plane with an associated symmetry axis, and show that each choice of this axis specifies a particular position operator, a corresponding position basis, and a position representation of the quantum mechanics of photon. In particular, we examine the position wave functions determined by such a position basis, elucidate their relationship with the Riemann-Silberstein and Landau-Peierls wave functions, and give an explicit formula for the probability density of the spatial localization of the photon.
\vspace{2mm}

\vspace{2mm}

\end{abstract}

\section{Introduction}

The question of the localizability of a photon in space has been one of the most basic and important questions of modern physics since the very inception of the notion of a photon. Although there is an extensive literature on the subject, a definitive and universally accepted answer to this question has not been available. This is mainly because a systematic and complete quantum mechanical treatment of a first-quantized photon could not be offered. The purpose of the present article is to provide such a treatment. Specifically we give an explicit construction of the following ingredients of the quantum mechanics of a photon.
    \begin{enumerate}
    \item A genuine Hilbert space of state vectors with a positive-definite and Lorentz-invariant inner 	product,
    \item Hermitian operators representing the Hamiltonian, momentum, helicity, and position observables,
    \item Localized states with definite helicity and sign of energy,
    \item Position wave functions and probability density for spatial localization of the photon.
    \end{enumerate}
The existing literature on the subject includes various attempts at constructing some of these quantities. For example, the study of the momentum observable and the momentum wave functions for a photon does not lead to any major difficulties \cite{1,2}. The opposite is the case when one tries to construct an appropriate position operator or directly define a position wave function.

The first major attempt at constructing a position wave function for a photon is due to Landau and Peierls (LP) \cite{3}. The result was found unsatisfactory, particularly by Pauli \cite{6}, who argued that the LP wave function did not transform like a tensor under Lorentz transformations and that it was a nonlocal function of the electric and magnetic fields, $\bfE$ and $\bfB$, of the photon. This in turn prevented its modulus square to serve as a measure of the probability of the interaction of the photon with localized charges \cite{6,1,7}. The LP wave function can be obtained by performing a nonlocal transformation on a local wave function \cite{1} such as the Riemann-Silberstein (RS) wave function, $\textbf{E}+i\textbf{B}$, \cite{19}. The latter has various useful properties \cite{1,7,18}, but its modulus (norm) square gives the energy density of the photon, not the probability density of its spatial localization.

The problem of finding an appropriate position wave function for a photon is clearly linked to the problem of constructing a position operator and localized states for the photon. A systematic study of relativistic position operators and localized states for particles of arbitrary spin is the subject of a seminal work of Newton and Wigner \cite{23} where the authors give a precise definition of a relativistic position operator and address the issues of its uniqueness and construction. This work generated a new wave of interest and led to mathematically more rigorous studies of the subject \cite{31}, but the results did not apply to photons \cite{23,31,33}. This motivated others to relax some of the stringent conditions on the notion of localizability and led to the development of a notion of weak localizability \cite{34} that turned out to be realizable for a photon (not having a definite helicity) \cite{35}. These developments, which rely on a number of highly technical mathematical results, avoid addressing the problem of obtaining explicit formulas for a sensible position operator for the photon (and the corresponding localized states and position wave functions) simply because their axiomatic basis implies the nonexistence of such an operator \cite{pike}. For a critical assessment of the nonexistence proof of the photon's position operator, see \cite{30}.

%As pointed out in Ref.~\cite{13}, the recent progress in the experimental studies of single photon states \cite{37} calls for a treatment of these problems.

The long list of contributions to the subject includes a pioneering work of Pryce \cite{24} which proposes an elaborate scheme for constructing relativistic position operators. This does actually lead to a position operator for a photon, but it is plagued by the fact that the components of this operator do not commute \cite{25}. This difficulty was to be circumvented more than half a century later in a work of Hawton \cite{11}, where she shows that adding an appropriate term to Pryce's position operator leads to a position operator with commuting components. The derivation of Hawton's position operator does not rely on the construction of a genuine Hilbert space for a photon. There is also a very limited information about the behavior of the corresponding localized states \cite{Hawton-1999b,30,HB,13} and position wave functions. In particular, the nature of the electric and magnetic field configurations for a localized photon is not known.

The formulation of the quantum mechanics of a photon that we offer in this article allows for a natural construction of a position operator with commuting components and yields explicit formulas for the corresponding localized states and position wave functions. It also elucidates the relationship between these and the previously obtained results such as the LP and RS wave functions and Hawton's position operator.

The conceptual framework for the present investigation is provided by the approach pursued in Refs.~\cite{cqg-2003,50,49,53,51} to formulate the quantum mechanics of Klein-Gordon and Proca
fields. This is in sharp contrast with axiomatic approach of Refs.~\cite{31,33,34,35}. In particular, rather than restricting ourselves by subscribing to certain axioms, we employ some very basic facts about quantum mechanics, inner products, and unitary operators to construct the building blocks of photon's quantum mechanics. To make our treatment self-contained we discuss these facts in the following.

First, we note that every quantum system is uniquely determined by a Hilbert space $\sH$ and a Hamiltonian operator acting in $\sH$, \cite{ps-2010}. The Hilbert space includes the state vectors of the system and fixes the set of Hermitian operators that represent its observables \cite{footnote1}. This together with the von Neumann's projection axiom determines the kinematic structure of the system. The dynamics is then defined by the Hamiltonian through the time-dependent Schr\"odinger equation.

By ``formulation of the quantum mechanics of a scalar or vector field'' we mean the identification of an appropriate Hilbert space and a Hamiltonian operator for the field. The natural candidates for these are respectively the space $\cV$ of solutions of the field equation(s) and the linear operator $h$ that generates time-translations in $\cV$. The key missing ingredient is an appropriate inner product on $\cV$ that makes it into a Hilbert space and  ensures the Hermiticity of $h$. Here by the qualification ``appropriate'' we mean that the inner product may be required to respect certain restrictions imposed by the physical/symmetry considerations. We may generally need to invoke the procedure of Cauchy completion to fulfill the mathematical requirement of the convergence of Cauchy sequences in $\sH$. We will however not delve into the details of such mathematical technicalities and pretend that what we know for finite-dimensions holds (or admits suitable generalizations) for the infinite dimensional spaces that we deal with \cite{footnote2}.

Next, we comment on a simple method of constructing inner products and specify the notation we employ throughout this article.

Consider a pair of vector spaces $\cX$ and $\cY$. Let $\br\cdot,\cdot\kt$ be an inner product on $\cY$ and $L:\cX\to\cY$ be a one-to-one linear operator defined on $\cX$. We can use $L$ and $\br\cdot,\cdot\kt$ to induce or pull back an inner product $\pbr\cdot,\cdot\pkt$ on $\cX$ according to $\pbr\psi,\phi\pkt:=\br L\psi,L\phi\kt$. Endowing $\cX$ with this inner product, we can identify $L$ with an isometry~\cite{footnote-2b} mapping $\cX$ onto its range, ${\rm Ran}(L)$. If $L$ is onto, ${\rm Ran}(L)=\cY$ and $L:\cX\to\cY$ is a unitary operator. In this case, we can express every Hermitian operator acting in $\cX$ in the form $L^{-1}OL$ where $O$ is a Hermitian operator acting in $\cY$.

We use the symbol $\bone_{n}$ to denote the $n\times n$ identity matrix, and employ the standard notation of labeling the Pauli and Gel-Mann matrices by $\bsigma_i$ and $\blambda_j$, respectively. In particular, we have
    \bea
    &\begin{aligned}
    &\bsigma_3:=\left[\begin{array}{cc} 1 & 0 \\  0 & -1
    \end{array}\right],~~~~~
    &&\lnd_2:=\left[\begin{array}{ccc}
        0 & -i & 0 \\ i & 0 & 0 \\ 0 & 0 & 0
      \end{array}\right],~~~~~
    &&\lnd_3:=\left[\begin{array}{ccc}
        1 & 0 & 0 \\ 0 & -1 & 0 \\ 0 & 0 & 0
      \end{array}\right],\nn\\[6pt]
    \end{aligned}&\\
    &\begin{aligned}
    &\lnd_5:=\left[\begin{array}{ccc}
        0 & 0 & -i \\ 0 & 0 & 0 \\ i & 0 & 0
      \end{array}\right],~~~~~
    &&\lnd_7:=\left[\begin{array}{ccc}
        0 & 0 & 0 \\ 0 & 0 & -i \\ 0 & i & 0
      \end{array}\right].
    \end{aligned}&
    \nn
    \eea
Clearly the standard basis of $\C^2$ and $\C^3$ consists of the eigenvectors of $\bsigma_3$ and  $\lnd_3$, respectively. We therefore label the basis vectors as follows.
    \bea
    &\begin{aligned}
    &\bbe_+:=\left[\begin{array}{c}1\\0\end{array}\right],~~~~~
    &&\bbe_-:=\left[\begin{array}{c}0\\1\end{array}\right],
    \end{aligned}&
    \label{0-2}\\
    &\begin{aligned}
    &\bbe_{1}:=\left[\begin{array}{c}1\\0\\0\end{array}\right],~~~~~
    &&\bbe_{-1}:=\left[\begin{array}{c}0\\1\\0\end{array}\right],~~~~~
    &&\bbe_{0}:=\left[\begin{array}{c}0\\0\\1\end{array}\right].
    \end{aligned}&
    \label{0-4}
    \eea
These satisfy  $\bsigma_3\bbe_{\epsilon}=\epsilon\,\bbe_{\epsilon}$ and $\lnd_3\bbe_s=s\,\bbe_s$, where $\epsilon=\pm$ and $s=-1,0,1$.

%Next, we introduce
%    \be
%    \begin{aligned}
%    &\tSigma_\fn:=\bsigma_i\otimes\bsigma_\ell,~~~~{\rm with}~~~~
%    \fn:=i+4\ell=0,1,\cdots,15,\\
%    &\bSigma_\fm:=\bsigma_i\otimes\lnd_j,~~~~{\rm with}~~~~
%    \fm:=i+4j=0,1,\cdots,35,
%    \end{aligned}
%    \label{0-5}
%    \ee
% where $i,\ell=0,1,2,3$ and $j=0,1,\cdots,8$. Then $\{\tSigma_\fn\}$ and $\{\bSigma_\fm\}$ respectively form a basis for the vector space of complex $4\times 4$ and $6\times 6$ matrices. It is easy to see that $\{\tSigma_0,\tSigma_3,\tSigma_{12},\tSigma_{15}\}$ and $\{\bSigma_0,\bSigma_3,\bSigma_{12},\bSigma_{15},\bSigma_{32},\bSigma_{35}\}$ are respectively maximal sets of commuting matrices with common eigenvectors
%    \begin{align}
%    &\tilde{\bbe}_{\ep,h}:=\bbe_\ep\otimes \bbe_h,
%    &&\bbe_{\ep,s}:=\bbe_\ep\otimes \bbe_s,
%    \label{0-6}
%    \end{align}
%where $\ep,h=\pm$ and $s=-1, 0,+1$.

Throughout this article we use $L^2(\R^3)$ for the space of square-integrable functions, $\xi:\R^3\to\C$, endowed with the inner product:
$\br\xi|\zeta\kt:=\int_{\R^3}d^3\bfx\, \xi(\bfx)^*\zeta(\bfx)$, where $\bfx=(x_1,x_2,x_3)$ represents the Cartesian coordinates of points in the Euclidean space $\R^3$, and an asterisk stands for complex-conjugation. Furthermore, we use $L^2(\R^3)\otimes\C^m$ to denote the space of vector fields $\bfxi:\R^3\to\C^m$ with square-integrable components $\xi_j$ endowed with the inner product $\bbr\cdot|\cdot\kkt$ defined
by
    \be
    \bbr\bfxi|\bfzeta\kkt=\sum_{j=1}^m\br\xi_j|\zeta_j\kt.
    \label{L2-m-dim}
    \ee
Because we can identify $L^2(\R^3)\otimes\C^m\otimes\C^n$ with $L^2(\R^3)\otimes\C^{mn}$, we use $\bbr\cdot|\cdot\kkt$ to label the inner product of $L^2(\R^3)\otimes\C^m\otimes\C^n$ as well.

The organization of this article is as follows. In order to familiarize the reader with our basic strategy, in Sec.~\ref{Sec2} we discuss a reformulation of nonrelativistic quantum mechanics of a free scalar field where we construct the Hilbert space using the space of solutions of the Sch\"odinger equation. In Sec.~\ref{Sec3} we employ this approach to outline the quantum mechanics of massive and massless relativistic free scalar fields. In Sec.~\ref{Sec4} we present our formulation of the quantum mechanics of a photon. Here we construct the Hilbert space and determine the observables of the photon. In particular, we obtain the operators representing photon's Hamiltonian, helicity, momentum, and position operators, and give explicit expressions for its localized states and position wave functions. In Sec.~\ref{Sec5} we summarize our findings and present our concluding remarks.

\section{Quantum Mechanics of a Nonrelativistic Free Scalar Field}
\label{Sec2}

\subsection{Space of fields and time-translations}
\label{Sec2A}

A nonrelativistic free complex scalar field $\psi:\R^4\to\C$ of mass $m$ is described by the field equation,
	\be
	i\partial_t\psi(\bfx,t)=-\frac{\hbar}{2m}\nabla^2\psi(\bfx,t),
	\label{sch-eq}
	\ee
whose solutions are uniquely determined by an initial condition of the form
	\be
	\psi(\bfx,t_0)=\psi_0(\bfx).
	\label{ini-condi}
	\ee
Here $t_0$ is a given initial time, and $\psi_0\in L^2(\R^3)$. This in turn implies that the function $\psi(t):\R^3\to\C$ defined by $\big(\psi(t)\big)(\bfx):=\psi(\bfx,t)$, is square-integrable for all $t\in\R$. We can therefore express (\ref{sch-eq}) and (\ref{ini-condi}) as the following initial-value problem in the Hilbert space $L^2(\R^3)$:
	\begin{align}
	& i\hbar\frac{d}{dt}\psi(t)=H_0\psi(t),
	\label{sch-eq-1}\\
	&\psi(t_0)=\psi_0,
	\label{ini-condi-1}
	\end{align}
where $H_0:L^2(\R^3)\to L^2(\R)$ is the linear operator given by
$(H_0\phi)(\bfx):=(-\hbar^2/2m)\nabla^2\phi(\bfx)$.

Next, we observe that every solution $\psi$ of (\ref{sch-eq-1}) defines a function
$\psi:\R\to L^2(\R^3)$ that maps $t$ to $\psi(t)$. In other words, we can identify every nonrelativistic free scalar field of mass $m$ with a solution $\psi:\R\to L^2(\R^3)$ of (\ref{sch-eq-1}), and express the set of all such fields as
	\be
	\cV:=\left\{\left. \psi:\R\to L^2(\R^3)~\right|~ \mbox{$i\hbar\frac{d}{dt}$}\psi(t)=H_0\psi(t)~\right\}.
	\label{sol-space-1}
	\ee	
This is clearly a complex vector space.

Let $\tau\in\R$, $\psi\in\cV$, and $\psi_\tau:\R\to L^2(\R^3)$ be defined by
	\be
	\psi_\tau(t):=\psi(t+\tau).
	\label{psi-tau-t-trans}
	\ee
Then it is easy to check that $\psi_\tau$ satisfies (\ref{sch-eq-1}), i.e., it belongs to $\cV$. This is the field obtained by translating $\psi$ in time by $\tau$. We can characterize the time-translations of the fields $\psi$ as linear operators  $u(\tau):\cV\to\cV$ defined by
	\be
	u(\tau)\psi :=\psi_\tau.
	\label{time-trans-1}
	\ee	

The existence and uniqueness of the solution of (\ref{sch-eq-1}) and (\ref{ini-condi-1}) means that each choice of $t_0$ specifies a one-to-one and onto function $V_{t_0}:L^2(\R^3)\to\cV$ according to $V_{t_0}\psi_0:=\psi$. Because  both $L^2(\R^3)$ and $\cV$ are complex vector spaces and (\ref{sch-eq-1}) is a linear equation, $V_{t_0}$ is a vector-space isomorphism (i.e., a one-to-one onto linear map.) The same holds for its inverse, $U_{t_0}:\cV\to L^2(\R^3)$, that satisfies
	\be
	U_{t_0}\psi=\psi(t_0).
	\label{U=def-1}
	\ee
In view of (\ref{psi-tau-t-trans}), (\ref{time-trans-1}), and (\ref{U=def-1}),
	\be
	U_{t_0}u(\tau)\psi=(u(\tau)\psi)(t_0)=\psi(t_0+\tau)=U_{t_0+\tau}\psi.
	\label{U-u-psi-1}
	\ee
This in turn implies that, for all $\tau,t_0\in\R$,
	\be
	u(\tau)=U_{t_0}^{-1}U_{t_0+\tau}.
	\label{small-u-tau=1}
	\ee

We can also obtain an explicit expression for the generator $h:\cV\to\cV$ of the time-translations, which by definition fulfills
	\be
	i\hbar \frac{d}{d\tau}u(\tau)=h u(\tau).
	\label{sch-eq-u-1}
	\ee
Applying both sides of this equation to an arbitrary $\psi\in\cV$ and making use of (\ref{U-u-psi-1}), (\ref{sch-eq-1}),  (\ref{U=def-1}), and (\ref{small-u-tau=1}), we have
	\bea
    i\hbar \frac{d}{d\tau}u(\tau)\psi
    &=&i\hbar\,U_{t_0}^{-1}\frac{d}{d\tau}\psi(t_0+\tau)
    %\nn\\&=&
    =U_{t_0}^{-1}H_0\psi(t_0+\tau)
    %\nn\\&=&
    =U_{t_0}^{-1}H_0 U_{t_0+\tau}\psi
    %\nn\\&=&
    =U_{t_0}^{-1}H_0 U_{t_0}u(\tau)\psi.\nn
    \eea
In light of (\ref{sch-eq-u-1}), this implies
	\be
	h=U_{t_0}^{-1}H_0 U_{t_0}.
	\label{h=U-D-D-1}
	\ee
	
Let us also note that we can use the isomorphism $U_{t_0}$ and the time-translation operator, $u(\tau):\cV\to\cV$, to induce a linear operator, $\cU(\tau):L^2(\R^3)\to L^2(\R^3)$, that maps each $\phi\in L^2(\R^3)$ to  $(V_{t_0}(\phi))(t_0+\tau)=U_{t_0+\tau}U_{t_0}^{-1}\phi$. It is easy to see that
	\be
	\cU(\tau)=U_{t_0+\tau}U_{t_0}^{-1}=U_{t_0}u(\tau)U_{t_0}^{-1}.
	\label{cU=1}
	\ee
The isomorphism $U_{t_0}$ provides an identical image of the time-translated field $\psi_\tau$ in $L^2(\R^3)$.  $\cU(\tau)$ is the time-translation operator in $L^2(\R^3)$ for this image. We can also determine the generator of time-translations in $L^2(\R^3)$, namely
	\be
	H:=i\hbar\left[\frac{d}{d\tau}\cU(\tau)\right]\cU(\tau)^{-1}.
	\label{Hamiltonian-1-big}
	\ee
Differentiating both sides of (\ref{cU=1}) with respect to $\tau$ and making use of (\ref{sch-eq-u-1}) -- (\ref{Hamiltonian-1-big}), we find
$H=H_0=-\frac{\hbar^2}{2m}\nabla^2$. Because $H$ does not depend on time, (\ref{Hamiltonian-1-big}) implies
	\be
	\cU(\tau)=e^{-\frac{i\tau}{\hbar} H}=e^{\frac{i \hbar\tau}{2m} \nabla^2}.
	\label{cU=exp-1}
	\ee
This in particular shows that $\cU(\tau)$ is a unitary operator acting in $L^2(\R^3)$.

\subsection{Hilbert space and Hamiltonian}
\label{Sec2B}

We identify the Hilbert space $\sH$ of our quantum system with the solution space $\cV$ endowed with such a positive-definite inner product, $\cbr\cdot,\cdot\ckt$, that the time-translations $u(\tau)$ act as unitary operators in $\sH$. This means that for all $\psi_1,\psi_2\in\sH$ and $\tau\in\R$,
$\cbr u(\tau)\psi_1,u(\tau)\psi_2\ckt=\cbr\psi_1,\psi_2\ckt$. We refer to every inner product $\cbr\cdot,\cdot\ckt$ fulfilling this relation as a ``dynamically invariant inner product'' or simply an ``invariant inner product.'' The unitarity condition is equivalent to the requirement that the generator $h$ of time-translations acts as a Hermitian operator in $\sH$. This makes it into the natural choice for the Hamiltonian operator of our quantum system.
	
The prescription we have outlined above reduces the construction of the quantum system of interest into the determination of an invariant inner product $\cbr\cdot,\cdot\ckt$ on $\cV$. We do this by pulling back $\cbr\cdot,\cdot\ckt$ from $L^2(\R^3)$ using the operator $U_{t_0}:\cV\to L^2(\R^3)$, i.e., set
	\be
	\cbr\psi_1,\psi_2\ckt:=\br U_{t_0}\psi_1|U_{t_0}\psi_2\kt=\br\psi_1(t_0)|\psi_2(t_0)\kt,
	\label{inner-product=1}
	\ee
where $\psi_1$ and $\psi_2$ are arbitrary elements of $\cV$, and $t_0$ is an arbitrary initial time. In view of Eqs.~(\ref{small-u-tau=1}), (\ref{cU=1}),  (\ref{cU=exp-1}), (\ref{inner-product=1}) and the fact that $\cU(\tau)$ acts as a unitary operator in $L^2(\R^3)$, we have
	\bea
	\cbr u(\tau)\psi_1,u(\tau)\psi_2\ckt&=&
	\cbr U_{t_0}^{-1}U_{t_0+\tau}\psi_1,U_{t_0}^{-1}U_{t_0+\tau}\psi_2\ckt
    %\nn\\&=&
    =\br U_{t_0+\tau}\psi_1|U_{t_0+\tau}\psi_2\kt\nn\\
    &=&\br\cU(\tau)U_{t_0}\psi_1|\cU(\tau)U_{t_0}\psi_2\kt
    %\nn\\&=&
    =\br U_{t_0}\psi_1|U_{t_0}\psi_2\kt
    %\nn\\&=&
    =\cbr \psi_1,\psi_2\ckt.
    \label{inv-3456}
	\eea
This shows that (\ref{inner-product=1}) is indeed an invariant inner product on $\cV$. We therefore identify $\sH$ with the Hilbert space obtained by giving this inner product to $\cV$.

We can view $U_{t_0}$ as a linear operator mapping $\sH$ onto $L^2(\R^3)$. According to
(\ref{inner-product=1}) this is a unitary operator. This observation together with  Eq.~(\ref{h=U-D-D-1}) and the Hermiticity of $H:L^2(\R^3)\to L^2(\R^3)$ imply that the Hamiltonian $h:\sH\to\sH$ is also a Hermitian operator. As a result $(\sH,h)$ defines a unitary quantum system.  The observables of this system are selected from among the Hermitian operators acting in $\sH$. These have the form $U_{t_0}^{-1}O\, U_{t_0}$, where $O$ is a Hermitian operator acting in $L^2(\R^3)$.

In view of Eq.~(\ref{h=U-D-D-1}) and the unitarity of $U_{t_0}:\sH\to L^2(\R^3)$ we can describe the quantum system given by $(\sH,h)$ in terms of $(L^2(\R^3),H_0)$. The latter provides the standard textbook treatment of a nonrelativistic free particle. We have taken the trouble of introducing the former representation of this system, because it admits an immediate generalization to relativistic fields.

\subsection{Position and momentum observables, localizes states, and position wave functions}
\label{Sec2C}

The operators $X_i,P_i:\sH\to\sH$ representing the components of the position and momentum observables of our quantum system are, by definition, required to satisfy the canonical commutation (Heisenberg algebra) relations:
	\begin{align}
	& [X_i,X_j]=[P_i,P_j]=0, && [X_i,P_j]=i\hbar \,\delta_{ij}1.
    	\label{H-algebra}
    	\end{align}
Here $i,j=1,2,3$, $\delta_{ij}$ stands for the Kronecker delta symbol, and $0$ and $1$ label the relevant zero and identity operators, respectively. We define $X_i$ and $P_i$ by
	\begin{align}
	&X_i:=U_{t_0}^{-1}\hat x_i U_{t_0},
	&&P_i:=U_{t_0}^{-1}\hat p_i U_{t_0},
	\label{X-P=1}
	\end{align}
where $\hat x_i$ and $\hat p_i$ are respectively the components of the standard position and momentum operators acting in $L^2(\R^3)$, i.e.,
	\begin{align}
	&(\hat x_i\phi)(\bfx):=x_i\phi(\bfx),
	&(\hat p_i\phi)(\bfx):=-i\hbar\frac{\partial}{\partial x_i}\phi(\bfx).
	\label{standard-x-p}
	\end{align}
	
Having determined the position operator, $\bfX:=(X_1,X_2,X_3)$, for our nonrelativistic scalar field, we introduce its localized state vectors according to
	\be
	\xi_{\bfy}:=U_{t_0}^{-1}\delta_{\bfy},
	\label{loc=1}
	\ee
where $\bfy$ is the point at which $\xi_{\bfy}$ is localized and $\delta_{\bfy}$ is the three-dimensional Dirac-delta function centered at $\bfy$, i.e.,
	\be
	\delta_{\bfy}(\bfx):=\delta^3(\bfx-\bfy)=\prod_{i=1}^3\delta(x_i-y_i).
	\label{delta=3}
	\ee
In view of (\ref{X-P=1}), (\ref{loc=1}), (\ref{delta=3}), and
    \be
    \hat x_i \delta_{\bfy}=y_i  \delta_{\bfy},
    \label{eg-va-x-1}
    \ee
$\xi_{\bfy}$ satisfies the eigenvalue equation $X_i \xi_{\bfy}=y_i \xi_{\bfy}$.

Next, we recall the following orthonormality and completeness properties of $\delta_{\bfy}$,
\cite{footnote3}.
	\bea
	&&\br\delta_{\bfy}|\delta_{\tilde{\bfy}}\kt=\delta^3(\bfy-\tilde\bfy),
	\label{delta-orthonormal}\\
	&&\phi=\int_{\R^3} d^3\bfy\, \br\delta_{\bfy}|\phi\kt \delta_{\bfy},
	\label{delta-completeness}
	\eea
where $\bfy,\tilde\bfy\in\R^3$ and $\phi\in L^2(\R^3)$. Equations (\ref{delta=3}) and (\ref{delta-completeness}) imply $\phi(\bfx)=\br\delta_{\bfx}|\phi\kt$. We can use (\ref{inner-product=1}), (\ref{loc=1}), (\ref{delta-orthonormal}), and (\ref{delta-completeness}) to derive the orthonormality and completeness relations for $\xi_{\bfy}$. These read
	\bea
	&&\cbr \xi_{\bfy},\xi_{\tilde\bfy}\ckt=\delta^3(\bfy-\tilde\bfy),
	\label{xi-orthonormal}\\
	&&\psi=\int_{\R^3} d^3\bfy\, \cbr \xi_{\bfy},\psi\ckt\,\xi_{\bfy}.
	\label{xi-completeness}
	\eea
	
We can interpret (\ref{xi-completeness}) as the expansion of the state vector $\psi$ in the ``position basis'' consisting of the localized state vectors $\xi_{\bfx}$. We identify the position wave function $f$ for $\psi$ with the function giving the coefficient of this expansion, i.e.,
	\be
	f(\bfx):=\cbr \xi_{\bfx},\psi\ckt.
	\label{position-wf-1}
	\ee	
Notice that in view of (\ref{inner-product=1}), (\ref{loc=1}), and (\ref{position-wf-1}),
$f(\bfx)=\br U_{t_0}\xi_{\bfx}|U_{t_0}\psi\kt=\br\delta_{\bfx}|\psi(t_0)\kt=\psi(\bfx,t_0)$.
This coincides with the standard expression for the position wave function at time $t_0$. Furthermore, we can use   (\ref{xi-orthonormal}), (\ref{xi-completeness}), and (\ref{position-wf-1}) to establish $\cbr\psi,\tilde\psi\ckt=\int_{\R^3}d^3\bfx\, f(\bfx)^*\tilde f(\bfx)=\br f|\tilde f\kt$, where $\psi$ and $\tilde\psi$ are any pair of elements of $\sH$, and $f$ and $\tilde f$ are the corresponding position wave functions.

According to von Neumann's projection axiom, the probability density $\rho(\bfx)$ for the spatial localization of the state given by $\psi\in\sH$ has the form
    \[ \rho(\bfx)=\frac{|\cbr\xi_{\bfx},\psi\ckt|^2}{\cbr\psi,\psi\ckt}=
    \frac{|f(\bfx)|^2}{\br f|f\kt}.\]
In the following sections we use this relation to identify the probability density of the spatial localization of a relativistic scalar field and a photon using the expression for the corresponding position wave functions. In order to determine the latter, we first need to construct the Hilbert space, position operator, and the localized states of the scalar field and the photon.

%We end this section by a remark on the choice of the localized state vectors $\delta_{\bfy}$ in the standard approach to quantum mechanics. It is important to notice that this choice follows from that of the standard position and momentum operators (\ref{standard-x-p}) which furnish a unitary irreducible representation of the Heisenberg algebra (\ref{H-algebra}). Although up to unitary equivalence this algebra admits a unique unitary irreducible representation. There are other unitary-equivalent representations that correspond to a different choice for the position and momentum operators. These in turn lead to different expressions for the localized state vectors than $\delta_{\bfy}$. For a particular example, see \cite{jpa-2006a}.

\section{Quantum Mechanics of a Relativistic free scalar field}
\label{Sec3}

The approach outlined in the preceding section can be consistently applied to develop a quantum mechanical treatment of relativistic free scalar fields. This is already done for the free massive scalar fields in Refs.~\cite{cqg-2003,49,50,53}. The treatment provided in these references, however, relies heavily on the properties of pseudo-Hermitian operators \cite{47,review}. In what follows we pursue an alternative approach that does not require a knowledge of these operators and allows for an explicit investigation of massless scalar fields.

Consider a complex scalar field $\psi:\R^4\to\C$ of mass $m$ that solves the field equation
    \be
    (\partial_0^2-\nabla^2+\fm^2)\psi(x^0,\bfx)=0,
    \label{field-eq-2}
    \ee
where $\partial_0$ stands for the differentiation with respect to $x^0:=ct$ and $\fm:=mc/\hbar$. Equation~(\ref{field-eq-2}) determines the field $\psi$ uniquely provided that we supplement it with a pair of initial conditions,
    \begin{align}
    &\psi(x_0^0,\bfx)=\psi_0(\bfx),
    &&\dot\psi(x_0^0,\bfx)=\chi_0(\bfx).
    \label{ini-condi-2}
    \end{align}
Here $x_0^0$ is an arbitrary initial value of $x^0$,  an over-dot stands for a derivative with respect to $x^0$, and $\psi_0,\chi_0\in L^2(\R^3)$.

\subsection{Hilbert space and Hamiltonian}
\label{Sec3A}

Equations (\ref{field-eq-2}) and (\ref{ini-condi-2}) define the following initial-value problem in $L^2(\R^3)$:
    \begin{align}
    &\ddot\psi(x^0)+D\psi(x^0)=0,
    \label{KG-eq-2}\\
    &\psi(x_0^0)=\psi_0,~~~~\dot\psi(x_0^0)=\chi_0,
    \label{init-condi-2a}
    \end{align}
where for each $x^0\in\R$, $\psi(x^0):\R^3\to\C$ is the function defined by $\big(\psi(x^0)\big)(\bfx):=\psi(x^0,\bfx)$, and $D:=-\nabla^2+\fm^2$.
%We can express the solution of (\ref{KG-eq-2}) and (\ref{init-condi-2a}) in the following implicit form \cite{50}.
%    \be
%    \psi(x^0)=\cos[(x^0-x^0_0)D^{1/2}]\psi_0+
%	\sin[(x^0-x^0_0)D^{1/2}]D^{-1/2}\chi_0.
%    \label{implicit}
%    \ee
Because $\psi_0$ and $\chi_0$ belong to $L^2(\R^3)$, the same holds for $\psi(x^0)$. We may therefore view the field $\psi$ as the function $\psi:\R\to L^2(\R^3)$ mapping $x^0$ to $\psi(x^0)$. In other words, we identify $\psi$ with an element of the complex vector space:
    \be
    \cV:=\left\{ \left. \psi:\R\to L^2(\R^3)~\right|~\ddot\psi(x^0)+D\psi(x^0)=0\right\}.
    \label{field-space-2}
    \ee
Our aim is to give this vector space the structure of a Hilbert space $\sH$ by endowing it with an invariant, positive-definite, and Lorentz-invariant inner product. To do this we introduce an analog of the isomorphisms $U_{t_0}$ of Sec.~\ref{Sec2}.

First, we recall that for each $\psi\in\cV$, the function $\psi_c:\R^4\to\C$ defined by
    \be
    \psi_c(x^0,\bfx):=\left(iD^{-1/2}\dot\psi(x^0)\right)(\bfx)
    \label{psi-c=}
    \ee
transforms as a Lorentz scalar \cite{53}.\footnote{Here and in what follows the powers of $D$ are defined in terms of its spectral representation, i.e., $(D^\alpha\phi)(\bfx):=(2\pi)^{-3/2}
\int_{\R^3} d^3\bfk\,(k^2+m^2)^\alpha e^{i\bfk\cdot\bfx}\tilde\phi(\bfk)$ where $\tilde\phi(\bfk):=
(2\pi)^{-3/2}\int_{\R^3} d^3\bfx\, e^{-i\bfk\cdot\bfx}\phi(\bfx)$ is the Fourier transform of $\phi(\bfx)$. This in turn implies that $(D^\alpha\phi)(\bfx)=\int_{\R^3}d^3\bfy\,\cK_\alpha(\bfx-\bfy)\phi(\bfy)$ where $\cK_\alpha(\bfx):=(2\pi)^{-3}\int_{\R^3} d^3\bfk\,(k^2+m^2)^\alpha e^{i\bfk\cdot\bfx}$.} It also satisfies the field equation (\ref{field-eq-2}). Therefore we can view it as an element of $\cV$. Next, we introduce
    \bea
    \Psi(x^0):=\frac{1}{2}\left[\begin{array}{cc}
    \psi(x^0)+\psi_c(x^0)\\
    \psi(x^0)-\psi_c(x^0)\end{array}\right],
    \label{two-component-2}
    \eea
which is an element of $\cH:=L^2(\R^3)\otimes\C^2$.
In view of (\ref{KG-eq-2}) it satisfies the Schr\"odinger equation,
    \be
    i\hbar\dot\Psi(x^0)={H}\Psi(x^0),
    \label{sch-eq-2}
    \ee
for the Hamiltonian operator:
    \be
    {H}:=\hbar\sqrt D\,\bsigma_3=\hbar\left[\begin{array}{cc}
    \sqrt D & 0\\
    0 & -\sqrt D\end{array}\right].
    \label{H-zero=2}
    \ee

Now, let $x_0^0\in\R$ and $U_{x_0^0}:\cV\to\cH$ be defined by
    \be
    U_{x_0^0} (\psi):=\Psi(x_0^0).
    \label{U-zeo-zero=2}
    \ee
It is not difficult to see that $U_{x_0^0}$ is an isomorphism of vector spaces. Using an analog of the analysis leading to (\ref {h=U-D-D-1}), we can show that the generator of time-translations in $\cV$ is
	\be
	h:=U_{x_0^0}^{-1}{H} U_{x_0^0}.
	\label{h=def-2}
	\ee
This observation together with the fact that ${H}$ acts as a Hermitian operator in $\cH$ shows that the inner product $\pbr\cdot,\cdot\pkt$ on $\cV$ that we obtain by pulling back the inner product of $\cH$ via $U_{x_0^0}$ is dynamically invariant.  It is not difficult to show that
    \bea
    \pbr\psi_1,\psi_2\pkt&:=&\bbr U_{x_0^0}\psi|U_{x_0^0}\psi\kkt
    =\frac{1}{2}\left[\br\psi_1(x_0^0)|\psi_2(x_0^0)\kt+\br\psi_{1c}(x_0^0)|\psi_{2c}(x_0^0)\kt\right]\nn\\
    &=&\frac{1}{2}\left[\br\psi_1(x_0^0)|\psi_2(x_0^0)\kt+
    \br\dot\psi_{1}(x_0^0)|D^{-1}\dot\psi_{2}(x_0^0)\kt\right],
    \label{innr-prod-2=}
    \eea
where $\bbr\cdot|\cdot\kkt$ denotes the inner product of $\cH$ that is given by (\ref{L2-m-dim}).

Evaluating the derivative of the right-hand side of (\ref{innr-prod-2=}) with respect to $x_0^0$ and using (\ref{KG-eq-2}) and (\ref{psi-c=}), we can check that indeed $\pbr\psi_1,\psi_2\pkt$ does not depend on the choice of $x_0^0$. It is also manifestly positive-definite, because $D^{-1}$ acts in $L^2(\R^3)$ as a positive-definite operator. The only problem is that $\pbr\psi_1,\psi_2\pkt$ fails to transform as a scalar under Lorentz transformations. This means that we should define the Hilbert space $\sH$ using a different inner product.

Let $V$ stand for the complex vector space consisting of the elements of $\cH$; so that
$\cH$ is ${V}$ endowed with the inner product $\bbr\cdot|\cdot\kkt$. We can construct the most general invariant inner product on $\cV$ (which makes $h$ Hermitian) by using $U_{x_0^0}$ to pullback the most general invariant inner product on ${V}$ (which makes ${H}$ Hermitian.) Because ${H}$ already acts as a Hermitian operator in $\cH$, a result of Ref.~\cite{jmp-2003} implies that every invariant inner product on ${V}$ has the form
	\be
	\bbr\cdot,\cdot\kkt_A:=\bbr A\cdot|A\cdot\kkt,
	\label{inn-prod-A-zero-2}
	\ee
where $A:\cH\to\cH$ is some invertible operator commuting with ${H}$. Pulling back $\bbr\cdot,\cdot\kkt_A$ to $\cV$ via $U_{x_0^0}$, we find the following invariant inner product on $\cV$.
    \be
    \pbr\psi_1,\psi_2\pkt_{\!\! A}:=\bbr U_{x_0^0}\psi_1,U_{x_0^0}\psi_2\kkt_A=
    \bbr A\,U_{x_0^0}\psi_1|A\,U_{x_0^0}\psi_2\kkt.
    \label{inn-prod-A-2}
    \ee

Following the approach of Ref.~\cite{cqg-2003}, we can show that imposing the requirement that
$\pbr\psi_1,\psi_2\pkt_{\!\! A}$ be Lorentz-invariant gives rise to a two-parameter family of inner products of the form:
	\bea
	\cbr\psi_1,\psi_2\ckt&=&\frac{\ell }{2}
	\Big\{\br\psi_1(x_0^0)|D^{1/2}\psi_2(x_0^0)\kt+
	\br\dot\psi_{1}(x_0^0)|D^{-1/2}\dot\psi_{2}(x_0^0)\kt+\nn\\
	&& i a\left[\br\psi_1(x_0^0)|\dot\psi_2(x_0^0)\kt
	-\br\dot\psi_1(x_0^0)|\psi_2(x_0^0)\kt\right]\Big\},
	\label{rel-inn-prod-2}
	\eea
where $\ell $ and $a$ are arbitrary real parameters, $\ell >0$, and $|a|<1$, \cite{footnote5}.
The values of these parameters are physically irrelevant, because $\ell $ drops out of the calculation of expectation values, and different choices of $a$ yield unitary-equivalent Hilbert space-Hamiltonian pairs. Therefore they determine the same physical system. Clearly $a=0$ is the simplest choice for $a$. Notice also that taking $\ell $ to have the dimension of length makes $\cbr\psi_1,\psi_2\ckt$ dimensionless. For a massive scalar field $\ell =1/\fm=$ Compton wavelength, is a natural choice \cite{cqg-2003,49,50}. In what follows we set $a=0$, but keep $\ell $ arbitrary, so that
	\be
	\cbr\psi_1,\psi_2\ckt=\frac{\ell }{2}
	\left[\br\psi_1(x_0^0)|D^{1/2}\psi_2(x_0^0)\kt+
	\br\dot\psi_{1}(x_0^0)|D^{-1/2}\dot\psi_{2}(x_0^0)\kt\right].
	\label{rel-inn-prod-inv-2}
	\ee
We define $\sH$ to be the Hilbert space obtained by endowing $\cV$ with this inner product.

In view of (\ref{innr-prod-2=}), (\ref{inn-prod-A-2}), and (\ref{rel-inn-prod-inv-2}), it is not difficult to see that $\cbr\psi_1,\psi_2\ckt=\pbr\psi_1,\psi_2\pkt_{\!\!A}$ for
$A=\sqrt\ell \,D^{1/4}\, {\bone_2}$. Furthermore, according to (\ref{inn-prod-A-2}), the relation
	\be
	U'_{x_0^0}:=AU_{x_0^0}=\sqrt\ell  D^{1/4}U_{x_0^0}
	\label{U-prime-2}
	\ee
defines a unitary operator mapping $\sH$ to $\cH$.

With the help of (\ref{h=def-2}), (\ref{H-zero=2}), and (\ref{U-prime-2}), we can show that
	\be
	U_{x_0^0}^{\prime-1}{H}U'_{x_0^0}=U_{x_0^0}^{-1}D^{-1/4}{H} D^{1/4}U_{x_0^0}=h.
	\label{h-UHU-prime-2}
	\ee
Because $U'_{x_0^0}:\sH\to\cH$ is a unitary operator and ${H}:\cH\to\cH$ is Hermitian, (\ref{h-UHU-prime-2}) identifies $h$ with a Hermitian operator acting in $\sH$. We define the quantum system for our relativistic scalar field by the pair $(\sH,h)$.

\subsection{Observables}
\label{Sec3B}

The observables of the quantum system determined by $(\sH,h)$ correspond to Hermitian operators acting in $\sH$. By virtue of the unitarity of $U'_{x_0^0}$, these  have the form $U_{x_0^0}^{\prime-1}OU'_{x_0^0}$ where $O$ is a Hermitian operator acting in $\cH$. The following are some basic examples.
	\bea
	C&:=&U_{x_0^0}^{\prime-1}\,\bsigma_3U'_{x_0^0}=
	U_{x_0^0}^{-1}\,\bsigma_3U_{x_0^0},
	\label{C=2}\\
	X_j&:=&U_{x_0^0}^{\prime-1}\hat x_j\,\bone_2\, U'_{x_0^0}=
	U_{x_0^0}^{-1}\cX_j U_{x_0^0},
	\label{Xj=2}\\
	P_j&:=&U_{x_0^0}^{\prime-1}\hat p_j\,\bone_2\, U'_{x_0^0}=
	U_{x_0^0}^{-1}\hat p_j\, U_{x_0^0},
	\label{Pj=2}
	\eea
where we have dropped the $2\times 2$ identity matrix $\bone_2$ whenever possible and introduced:
    \be
    \cX_j:=\hat x_j+\frac{i\hbar \hat p_j}{2(\hat\bfp^2+m^2c^2)},~~~~~~~~\hat \bfp^2:=\sum_{j=1}^3\hat p_j^2.
    \ee

$C$ is an involution ($C^2=1$) having eigenvalues $\epsilon=\pm 1$ and commuting with $h$. Therefore it is a symmetry generator that splits the Hilbert space $\sH$ into an orthogonal direct sum of its eigenspaces $\sH_\epsilon$. These contain scalar fields with a definite energy of sign $\epsilon$ and their superpositions. Ref.~\cite{53} establishes the remarkable identity,
    \be
    C\psi=\psi_c,
    \label{C=psi-c}
    \ee
and  uses the term ``chirality operator'' to refer to $C$. This operator has an analog in quantum cosmology that corresponds to a ``direction-of-time observable'' \cite{49}. This interpretation is also applicable in the present context. To see this let $\psi^\epsilon$ be any element of $\sH_\epsilon$, so that $\psi^\epsilon_c=C\,\psi^\epsilon=\epsilon\,\psi^\epsilon$.
Combining this relation with (\ref{psi-c=}) and (\ref{C=psi-c}), we find $\dot\psi^\epsilon(x^0)=-i\epsilon D^{1/2}\psi^\epsilon(x^0)$. This in turn implies
    \be
    \psi^\epsilon(x^0)= e^{-i\epsilon (x^0-x^0_0) D^{1/2}}\psi^\epsilon(x_0^0).
    \label{psi-ep-time}
    \ee
Therefore $\psi^-$ evolves backward in time (if $\psi^+$ evolves in forward direction in time.)

We can use the chirality and momentum operators, $C$ and $\bfP:=(P_1,P_2,P_3)$, to give a closed form expression for the generator of time translations in $\sH$, namely
    \be
	h=\left(\bfP^2+m^2c^2\right)^{1/2}C.
	\label{h=PC}
	\ee
This relation is a consequence of (\ref{H-zero=2}), (\ref{h-UHU-prime-2}), (\ref{C=2}), and (\ref{Pj=2}).

$X_i$ and $P_i$ furnish a unitary irreducible representation of the Heisenberg algebra (\ref{H-algebra}). We therefore identify them with the components of the position and momentum operators for our relativistic scalar field. We would like to stress that being linear operators mapping scalar fields to scalar fields, $C, X_i$, and $P_i$ are, by construction, Lorentz invariant.

Ref.~\cite{53} gives explicit formulas for the action of $X_j$ and $P_j$ on a scalar field $\psi$. They read
    \bea
    (X_j\psi)(x^0,\bfx)&=&\big(\fX_j\psi(x^0)\big)(\bfx),\nn\\
    (P_j\psi)(x^0,\bfx)&=&\big(\hat p_j\psi(x^0)\big)(\bfx),
    \eea
where
    \be
    \fX_j:=\hat x_j+\frac{i\hbar \hat p_j}{2(\hat\bfp^2+m^2c^2)}-\frac{i\hbar(x^0-x^0_0)\hat p_j}{\hat\bfp^2+m^2c^2}\,\partial_0.
    \ee
As noted in Ref.~\cite{50} this coincides with the components of the Newton-Wigner position operator \cite{23} provided that we set $x^0=x^0_0$ and restrict our attention to positive-energy scalar fields.

\subsection{Localized states and position representation}
\label{Sec3C}

Because $[X_j,C]=0$, we can construct scalar fields $\psi^{\ep}_{\bfy}$ that are localized at a point $\bfy$ in space and have a definite sign of energy $\epsilon$. These satisfy
    \begin{align*}
    & X_j\psi^{\ep}_{\bfy}=y_j \psi^{\ep}_{\bfy},
    && C\psi^{\ep}_{\bfy}=\epsilon\, \psi^{\ep}_{\bfy}.
    \end{align*}
In view of (\ref{eg-va-x-1}), (\ref{C=2}), and (\ref{Xj=2}), we can construct $\psi^{\ep}_{\bfy}$ by pulling back $\delta_{\bfy}\bbe_\epsilon$ via $U'_{x_0^0}$, i.e., setting
    \be
    \psi^{\ep}_{\bfy}:=U_{x_0^0}^{\prime -1} \delta_{\bfy}\bbe_\epsilon=
    \ell ^{-1/2} U_{x_0^0}^{-1} D^{-1/2} \delta_{\bfy}\bbe_\epsilon,
    \label{loc-psi-2}
    \ee
where $\bbe_\epsilon$ are the standard basis vectors of $\C^2$ that we give in (\ref{0-2}). We can use the unitarity of $U'_{x_0^0}$ and the orthonormality and completeness of $\delta_{\bfy}\bbe_\epsilon$ in $\cH$, namely
	\begin{align*}
	&\bbr \delta_{\bfy}\bbe_\epsilon, \delta_{\tilde\bfy}\bbe_{\tilde\epsilon}\kkt=
	\delta_{\epsilon \tilde\epsilon}\,\delta(\bfy-\tilde\bfy),
	&&
	\boldsymbol{\phi}=\sum_{\epsilon=\pm}\int_{\R^3} d^3\bfy\,
	\bbr\delta_{\bfy}\bbe_\epsilon,\boldsymbol{\phi}\kkt\, \delta_{\bfy}\bbe_\epsilon,
	\end{align*}
to establish the following orthonormality and completeness relations for $\psi^{\ep}_{\bfy}$.
	\begin{align}
	&	\cbr \psi^{\ep}_{\bfy},\psi^{\tilde\ep}_{\tilde{\bfy}}\ckt=
	\delta_{\epsilon \tilde\epsilon}\, \delta^3(\bfy-\tilde\bfy),\\
	&	\psi=\sum_{\epsilon=\pm}\int_{\R^3} d^3\bfy\,
	\cbr\psi^\epsilon_{\bfy},\psi\ckt\, \psi^\epsilon_{\bfy}.
	\label{psi-completeness-2}
	\end{align}

In order to determine the explicit form of the localized fields $\psi^{\ep}_{\bfy}(x^0,\bfx)$, we first introduce a basis consisting of the common eigenvectors of $C$ and $P_j$. Let
	\be
	\varphi_{\bfk}^\epsilon:=U^{\prime-1}_{x_0^0}(\phi_{\bfk}\bbe_\epsilon),
	\label{varphi=2a}
	\ee
where $\bfk\in\R^3$ and
    \be
    \phi_\bfk(\bfx):=(2\pi)^{-3/2}e^{i\bfk\cdot\bfx}.
    \label{phi=3}
    \ee
Then, as a result of (\ref{C=2}), (\ref{Pj=2}), and (\ref{h=PC}), we have
	\begin{align}
	&C\varphi_{\bfk}^\epsilon=\epsilon\,\varphi^\epsilon_{\bfk},
	&& P_j\varphi_{\bfk}^\epsilon=\hbar k_j\varphi^\epsilon_{\bfk},
	&& h\,\varphi_{\bfk}^\epsilon=\epsilon\,\hbar\sqrt{k^2+\fm^2}\,\varphi^\epsilon_{\bfk},
	\label{CPh-varphi}
	\end{align}
where $k:=\sqrt{\bfk^2}$. We can use (\ref{U-zeo-zero=2}), (\ref{U-prime-2}), and (\ref{varphi=2a}) to show that $\varphi_{\bfk}^\epsilon(x^0_0)=\ell ^{-1/2}D^{-1/4}\phi_{\bfk}$. This together with (\ref{psi-ep-time}) imply
	\be
	\varphi_{\bfk}^\epsilon(x^0,\bfx)=
	\frac{e^{-i\epsilon (x^0-x^0_0) \sqrt{k^2+\fm^2}}\phi_{\bfk}(\bfx)}{\sqrt\ell \: (k^2+\fm^2)^{1/4}}.
	\label{varphi=2}
	\ee

Next, we note that because $\{\phi_{\bfx}\bbe_\epsilon\}$ is an orthonormal basis for $\cH$ and $U'_{x^0_0}:\sH\to\cH$ is a unitary operator,
$\varphi_{\bfk}^\epsilon$ satisfy the orthonormality and completeness relations
    \be
    \cbr \varphi_{\bfk}^\epsilon,\varphi_{\tilde\bfk}^{\tilde\epsilon}\ckt=
    \delta_{\epsilon \tilde\epsilon}\,\delta^3(\bfk-\tilde\bfk),~~~~~~
    \sum_{\epsilon=\pm}\int_{\R^3}d^3\bfk \: \cbr\varphi^\epsilon_{\bfk},\psi\ckt\,\varphi^	 \epsilon_{\bfk}=\psi.
    \ee
Applying the latter formula for $\psi=\psi^\epsilon_{\bfy}$, noting that
$\cbr\varphi^{\tilde\epsilon}_{\bfk},\psi^\epsilon_{\bfy}\ckt=
	\bbr \phi_{\bf k}\bbe_{\tilde\epsilon}|\delta_{\bfy}\bbe_\epsilon\kkt=
	\br\phi_{\bfk}|\delta_{\bfy}\kt\,\bbe_{\tilde\epsilon}^*\cdot\bbe_\epsilon=
	\phi_{\bfk}(\bfy)^*\delta_{\tilde\epsilon \epsilon}
$, and making use of (\ref{phi=3}) and (\ref{varphi=2}), we obtain
	\bea
	\psi^\epsilon_{\bfy}(x^0,\bfx)&=&
	\int_{\R^3}\!\!\!d^3\bfk \: \phi_{\bfk}(\bfy)^*\varphi^{\tilde\epsilon}_{\bfk}(x^0,\bfx)
    =\frac{1}{2\pi^2\sqrt\ell \,|\bfx-\bfy|}\int_0^\infty\!\!\! dk\: \frac{k\sin(|\bfx-\bfy|k)
	e^{-i\epsilon(x^0-x^0_0) \sqrt{k^2+\fm^2}}}{(k^2+\fm^2)^{1/4}}.~~~~~~~
	\label{LF=int-2}
	\eea
We explore the consequences of this relation for the cases of massive and massless scalar fields separately.
	
For $\fm\neq 0$, we could evaluate the integral appearing in (\ref{LF=int-2}) only for $x^0=x^0_0$. This gives
	\be
    	\psi^{\ep}_{\bfy}(x_0^0,\bfx)=\frac{\alpha_0}{\sqrt\ell }  \left(\f{\fm}
        {|\bfx-\bfy|}\right)^{\f{5}{4}}K_{\f{5}{4}}(\fm |\bfx-\bfy|),
        \label{LS-KG}
    	\ee
where $\alpha_0:=[2^{3/4}\pi^{3/2}\Gamma(1/4)]^{-1}$, and $\Gamma(x)$ and $K_\nu(x)$ are respectively the Euler Gamma function and the Bessel K-function  \cite{53}. Equation~(\ref{LS-KG}) coincides with the formula obtained by Newton and Wigner for the localized Klein-Gordon fields in \cite{23}. In order to obtain $\psi^{\ep}_{\bfy}(x_0^0,\bfx)$ for $x^0\neq x^0_0$, we make use of the fact that $\psi^{\ep}_{\bfy}$ is an eigenvector of $C$ with eigenvalue $\epsilon$. This allows us to employ (\ref{psi-ep-time}) which gives
	\be
    \psi^{\ep}_{\bfy}(x^0,\bfx)=
	\left(e^{-i\epsilon (x^0-x^0_0)(\hat k^2+\fm^2)^{1/2}}\psi^{\ep}_{\bfy}(x_0^0)\right)(\bfx).
    \label{eq-add1}
    \ee

For $\fm=0$, the integration of the right-hand side of (\ref{LF=int-2}) gives
    \bea
    \psi^{\ep}_{\bfy}(x^0,\bfx)&=&\frac{1}{4(2\pi)^{3/2}\sqrt{\ell }\:|\bfx-\bfy|}\:
    \sum_{\gamma=\pm}\frac{1+i\gamma\epsilon}{\left[|\bfx-\bfy|-\gamma(x^0-x_0^0)\right]^{3/2}}.
    \label{LS-KG-zero}
    \eea
In particular,
    \be
    \psi^{\ep}_{\bfy}(x_0^0,\bfx)=\f{\pi}{ \sqrt\ell  \big[2\pi |\bfx-\bfy|\big]^{5/2}}.
    \label{LS-KG-zero-zero}
    \ee
We have checked that this result coincides with the $\fm\to0$ limit of (\ref{LS-KG}), as originally envisaged by Newton and Wigner \cite{23}.

Having obtained the localized states of our quantum system, we can identify the position wave function for a field $\psi\in\sH$ with the coefficient of its expansion in the localized state vectors $\psi^\epsilon_{\bfx}$. In light of (\ref{psi-completeness-2}), this has the form
	\bea
	f(\epsilon,\bfx)&:=&\cbr \psi_{\bfx}^\epsilon,\psi\ckt=
	\bbr U'_{x_0^0}\psi_{\bfx}^\epsilon|U'_{x_0^0}\psi\kkt
    %\nn\\&=&
	=\sqrt\ell \,\bbr \delta_{\bfx}\bbe_\epsilon|D^{1/4}\Psi(x_0^0)\kkt\nn\\
	&=&
	\frac{\sqrt\ell }{2}\br\delta_{\bfx}|D^{1/4}[\psi(x_0^0)+\epsilon\,\psi_c(x_0^0)]\kt
    %\nn\\&=&
	=\frac{\sqrt\ell }{2}\left[D^{1/4}\psi(x_0^0,\bfx)+
	i\epsilon\,D^{-1/4}\dot\psi(x_0^0,\bfx)\right].
	\eea
Let $\tilde f(\epsilon,\bfx)$ be the position wave function for $\tilde\psi\in\sH$. Then (\ref{psi-completeness-2}) implies that
	\bea
	\cbr \tilde\psi,\psi\ckt&=&
	\sum_{\epsilon=\pm}\int_{\R^3}d^3\bfx\, \cbr\psi^\epsilon_{\bfx},\psi\ckt
	\cbr\tilde\psi,\psi^\epsilon_{\bfx}\ckt
    %\nn\\&=&
    =\sum_{\epsilon=\pm}\int_{\R^3}d^3\bfx\,\tilde f(\epsilon,\bfx)^*f(\epsilon,\bfx)
	=\sum_{\epsilon=\pm}\br \tilde f(\epsilon,\cdot)|f(\epsilon,\cdot)\kt.~~~
	\eea
The probability density of the spatial localization of a field $\psi$ is given by
	\bea
	\rho(x_0^0,\bfx)&=&\sum_{\epsilon=\pm}\frac{|\cbr \psi_{\bfx}^\epsilon,\psi\ckt|^2}{
	\cbr\psi,\psi\ckt}
	=\frac{\sum_{\epsilon=\pm}|f(\epsilon,\bfx)|^2}{
	\sum_{\epsilon=\pm}\br f(\epsilon,\cdot)|f(\epsilon,\cdot)\kt}
    %\nn\\&=&
    =\frac{|D^{1/4}\psi(x_0^0,\bfx)|^2+|D^{-1/4}\dot\psi(x_0^0,\bfx)|^2}
	{\br \psi(x_0^0)|D^{1/2}\psi(x_0^0)\kt+
	\br \dot\psi(x_0^0)|D^{-1/2}\dot\psi(x_0^0)\kt}.~~~~
    \label{prob-dens-1}
	\eea
This is a function with nonnegative real values. Although it fails to be the time-like component of a conserved $4$-current, its integral over $\R^3$ coincides with unity. Therefore the total probability is indeed conserved. This turns out to be related to the local conservation of a complex current density \cite{53}.

Having obtained the position wave function for the states of a scalar field, we can describe both the kinematics and dynamics of quantum mechanics of a scalar field using its position representation. For a massive scalar field this is explained in Ref.~\cite{50}. The same approach applies to a massless scalar field. Here we summarize it for completeness: The position wave functions $f(\epsilon,\bfx)$ define elements $f$ of $L^2(\R^3)\oplus L^2(\R^3)$, because
    \[\sum_{\epsilon_\pm}\int_{\R^3}d^3\bfx|f(\epsilon,\bfx)|^2=\cbr\psi,\psi\ckt <\infty.\]
To each observable $O:\sH\to\sH$ and $\psi\in\sH$, we can associate a Hermitian operator $o:L^2(\R^3)\oplus L^2(\R^3)\to L^2(\R^3)\oplus L^2(\R^3)$ that maps elements $f$ of $L^2(\R^3)\oplus L^2(\R^3)$ to the position wave function for the state vector $O\psi$ where $\psi$ is the state whose position wave function is $f(\epsilon,\bfx)$. The operator $o$ is the position representation of $O$. We can use the former to describe the observable given by the latter. Similarly we can formulate the dynamics using position wave functions. It is easy to show that the position wave function $f(\epsilon,\bfx;x^0)$ for an evolving state vector $e^{-i(x^0-x^0_0)h/\hbar}\psi$ satisfies the
Schr\"odinger equation \cite{50}:
    \[ i\hbar\partial_0 f(\epsilon,\bfx;x^0)=\epsilon\sqrt{-\hbar^2\nabla^2+m^2c^2}f(\epsilon,\bfx;x^0).\]

\subsection{Lorentz transformation of localized states}
\label{Sec3D}

The term ``localized state'' has been used in the literature for different purposes. Sometimes it means a field configuration whose charge or energy density vanishes outside a small compact subset $V$ of a space-like hypersurface that is given by $x^0=y^0_0$ in some inertial coordinate frame, or that it decays rapidly as $|\bfx-\bfy|\to\infty$ for some $\bfy\in V$, \cite{BB-1998}.  There are well-known no-go theorems that define ``localization'' by requiring that certain ``basic'' and ``natural'' assumptions hold and then use these to establish its nonexistence \cite{hegerfeldt}. As noted in \cite{barat} one must exercise extra care in the use of these results because it may turn out that the assumptions they rely on are not realizable.

Our definition of a localized state is based solely on the measurement (projection) axiom of quantum mechanics applied to a position measurement. Consider an inertial observer $\cO$ who uses $x:=(x^0,\bfx)$ to label spacetime points, and suppose that she makes a simultaneous measurement of the sign of energy and position of a free scalar particle at a time $y_0^0$ when it is in the state given by $\psi\in\sH$. If the outcome of this measurement is $\eta$ for the sign of energy and $\bfy$ for the position, then the act of measurement projects the state onto the localized state given by $\psi^{\eta}_{\bfy}$. We call $\bfy$ the localization center of $\psi^{\eta}_{\bfy}$, because the position wave function for $\psi^{\eta}_{\bfy}$ has the form $f(\bfx,\ep)=\delta_{\ep\eta}\delta(\bfx-\bfy)$.
The probability that $\bfy$ lies in a region $V$ of the hyperspace $x^0=y^0_0$ is $\int_{V}d^3\bfy \rho(y_0^0,\bfy)$.

As seen from (\ref{prob-dens-1}), the variable $\bfy$ appearing in the expression for the probability density $\rho(y_0^0,\bfy)$ is the label identifying the localized state vector $\psi^{\eta}_{\bfy}$. Because this is by construction a scalar field, under a proper orthochronous Poincar\' e transformation, $x\to x'=\Lambda x+a$, it transforms according to
    \be
    \psi^{\eta}_{\bfy}(x)
    \xrightarrow{\makebox[1cm]{\small $(\Lambda,a)$}}
    \psi^{\prime\eta}_{\:\bfy}(x'):=\psi^{\eta}_{\bfy}(\Lambda^{-1}(x'-a)).
    \nn
    \ee
If we view this transformation as a change of coordinates to those used by an inertial observer $\cO'$, then the transformed field $\psi^{\prime\eta}_{\:\bfy}$ is the state vector that $\cO'$ uses to describe the localized state with sign of energy $\eta$ and center $\bfy$ in the frame of $\cO$. This is not generally the same as a localized state in the frame of $\cO'$, because the latter would be centered at a point on the space-like hypersurface $x^{\prime 0}=y_0^{\prime 0}$ for some $y_0^{\prime 0}$.

The standard quantum measurement theory applied to position measurements requires a splitting
of spacetime into space+time and a preferred time at which the observer makes the measurement.  All inertial observers must however agree on the theoretical predictions associated with a position measurement made by any one of them. In this connection, it is worthy of noting that they would all agree on the expression for the components of the position operator $X_i$, position wave function $f(\ep,\bfx)$, and the probability density $\rho(y_0^0,\bfy)$ associated with the frame of $\cO$, because these quantities are Lorentz-invariant  \cite{footnote65}.

If we view the probability density $\rho$ as a function mapping spacetime points $(y_0^0,\bfy)$ to real numbers, then we can show that it is not the time component of a four-vector. This is often viewed as a serious deficiency and used to argue for the nonexistence of position operator and localized states. By virtue of the above-mentioned frame-dependence of position measurements the above covariance requirement on $\rho$ is by no means justified.\footnote{The standard textbook proof of the identification of the probability density (of nonrelativistic QM) with the time component of a four-vector applies only for Hamiltonians that are quadratic polynomials in momenta. For example it fails for a Hamiltonian of the form $H=p^4+v(x)$, whose use is not prohibited by any of the axioms of quantum mechanics.}  As noted in \cite{currie-1963}, the principle of relativity only demands that the laws of nature have the same form in all inertial frames. The requirement that the quantities involved in the mathematical expression of these laws should be covariant does not follow from the basic axioms of special relativity or quantum mechanics \cite{35}. The formalism we have developed in this article provides a computational scheme for finding the probability of the outcome of measurements done by an inertial observer $\cO$. This scheme is applicable in all inertial frames and all inertial observers who use it to compute the probability of the outcome of a measurement performed by $\cO$ will find the same value. In this sense our scheme complies with the requirement of relativistic covariance.

\section{Quantum Mechanics of a Free Photon}
\label{Sec4}

In this section we develop an extension of the analysis of Sec.~\ref{Sec3} that applies to a first quantized free photon. We identify the latter with a massless complex vector field $A=(A^0,\bfA)$ whose field strength satisfies Maxwell's equations in vacuum \cite{Jackson}. In Gaussian units these have the form
    \begin{align}
    &\bnabla\cdot\bfE(x^0,\bfx)=0, && \partial_0\bfE(x^0,\bfx)=\bnabla\times\bfB(x^0,\bfx),
    \label{maxwell-3}
    \end{align}
where $\bfE$ and $\bfB$ respectively stand for the electric and magnetic fields associated with $A$, i.e.,
    \begin{align}
    &\bfE(x^0,\bfx):=-\partial_0\bfA(x^0,\bfx)-\bnabla A^0(x^0,\bfx), \\
    &\bfB(x^0,\bfx):=\bnabla\times\bfA(x^0,\bfx).
    \label{E-B=def}
    \end{align}

It is well-known that Maxwell's equations in vacuum are consistent with simultaneous imposition of the temporal and Coulomb gauge conditions \cite{Jackson},
    \begin{align}
    &A^0(x)=0, &&\bnabla\cdot\bfA(x)=0.
    \label{TC-gauge}
    \end{align}
In what follows we adopt the temporal-Coulomb gauge where (\ref{TC-gauge}) holds and
    \be
    \bfE(x^0,\bfx)=-\partial_0\bfA(x^0,\bfx).
    \label{E=TC-gauge}
    \ee
With the help of this relation and (\ref{E-B=def}) we can write Maxwell's equations~(\ref{maxwell-3}) in the form
    \bea
    \bnabla\cdot\partial_0\bfA(x^0,\bfx)&=&0,
    \label{constract-A-dot}\\
    (\partial_0^2-\nabla^2)\bfA (x^0,\bfx)&=&0.
    \label{dyn-eq-3}
    \eea
The latter is a wave equation for $\bfA (x^0,\bfx)$ that is to be solved under the constraints provided by the second relation in (\ref{TC-gauge}) and Eq.~(\ref{constract-A-dot}). The solution exists and is unique provided that we supplement (\ref{dyn-eq-3}) with a pair of initial conditions of the form:
    \begin{align}
    &\bfA (x_0^0,\bfx)=\bfA_0(\bfx),
    &&\partial_0\bfA (x_0^0,\bfx)=-\bfE_0(\bfx),
    \label{ini-condi-3}
    \end{align}
where $\bfA_0$ and $\bfE_0$ are a pair of elements of $\tilde\cH:=L^2(\R^3)\otimes\C^3$ that fulfil
    \be
    \bnabla\cdot \bfA_0(\bfx)=\bnabla\cdot \bfE_0(\bfx)=0.
    \label{ini-condi-constraint}
    \ee
In other words it suffices to impose the constraints on the initial data. It is the presence of these constraints that complicates the construction of the Hilbert space and observables for a photon as compared to its massive cousin, the Proca field \cite{51}.

\subsection{Hilbert space}
\label{Sec4A}

Following the approach of Secs.~\ref{Sec2} and \ref{Sec3}, for each $x^0\in\R$ we use $\bfA:\R^4\to\C^3$ to define a function $\bfA(x^0):\R^3\to\tilde\cH$ according to $\big(\bfA(x^0)\big)(\bfx):=\bfA(x^0,\bfx)$. Let us also introduce $\hat k_j,D,\hat k:L^2(\R^3)\to L^2(\R^3)$ and $\hat\bfk:L^2(\R^3)\to \tilde\cH$, as the operators defined by $(\hat k_i\phi)(\bfx):=-i\partial_j\phi(\bfx)$, $(D\phi)(\bfx):=-\nabla^2\phi(\bfx)$,
%	\begin{align*}
%	&(\hat k_i\phi)(\bfx):=-i\partial_j\phi(\bfx),
%	&& (D\phi)(\bfx):=-\nabla^2\phi(\bfx),
%	\end{align*}
$\hat\bfk:=(\hat k_1,\hat k_2,\hat k_3)$, and $\hat k:=\sqrt{\hat\bfk^2}$, so that $(\hat\bfk\phi)(\bfx)=-i\bnabla\phi(\bfx)$, $\hat\bfk^2=D$, and $\hat k=\sqrt D$.
%	\begin{align*}
%	&(\hat\bfk\phi)(\bfx)=-i\bnabla\phi(\bfx),
%	&& \hat\bfk^2=D,
%	&& \hat k=\sqrt D.
%	\end{align*}
Then we can respectively express the wave equation (\ref{dyn-eq-3}),  the initial conditions (\ref{ini-condi-3}), and the constraints (\ref{ini-condi-constraint}) as
    \begin{align}
    &\ddot\bfA(x^0)+D\bfA(x^0)=0,
    \label{wave-eqn-tH-3}\\
    &\bfA(x_0^0)=\bfA_0,~~~~\dot\bfA(x_0^0)=-\bfE_0,
    \label{ini-condi-3n}\\
    &\hat\bfk\cdot\bfA(x_0^0)=\hat\bfk\cdot\dot\bfA(x_0^0)=0.
    \label{constraints}
    \end{align}

Each complex vector field $A:\R^4\to\C^4$ that describes a photon in the temporal-Coulomb gauge defines a function $\bfA:\R\to\tilde\cH$ satisfying (\ref{wave-eqn-tH-3}) and (\ref{constraints}). We therefore identify the state vectors of the photon with the elements of the complex vector space:
    \be
    \cV:=\left\{\bfA:\R\to\tilde\cH \left|\begin{array}{c}
    \ddot\bfA(x^0)+D\bfA(x^0)=0~\mbox{for all}~ x^0\in\R,~{\rm and}\\
    \hat\bfk\cdot\bfA(x_0^0)=\hat\bfk\cdot\dot\bfA(x_0^0)=0
    ~\mbox{for some}~ x_0^0\in\R.\end{array}\right.\right\}.
    \ee
We wish to endow this vector space with a positive-definite inner product that is invariant under time translations:  $\bfA\to{\bfA}_{\tau}$, where ${\bfA}_{\tau}(x^0):=\bfA_{\tau}(x^0+c\tau)$.

First, we note that for each $\bfA\in\cV$, the relation
    \be
    \bfA_c(x^0):=iD^{-1/2}\dot\bfA(x^0)=i\hat k^{-1}\dot\bfA(x^0)
    \label{A-c=def}
    \ee
defines an element $\bfA_c$ of $\cV$, \cite{footnote6}. We use $\bfA$ and $\bfA_c$ to define the six-component vector:
    \be
    \Psi(x^0):=\frac{1}{2}\left[\begin{array}{c}
    \bfA(x^0)+\bfA_c(x^0)\\
    \bfA(x^0)-\bfA_c(x^0)\end{array}\right],
    \label{6-comp-3}
    \ee
that is an element of $\tilde\cH\otimes\C^2= L^2(\R^3)\otimes\C^3\otimes\C^2$. For brevity we use $\cH$ to label this Hilbert space. In terms of $\Psi(x^0)$ the constraint (\ref{constraints}) takes the form
    \be
    \hat\bfk\cdot\Psi(x_0^0)=0.
    \label{constraints-Psi}
    \ee
Furthermore, we can use (\ref{A-c=def}) and (\ref{6-comp-3}) to express the wave equation (\ref{wave-eqn-tH-3}) as the Schr\"odinger equation (\ref{sch-eq-2}) for the Hamiltonian
    \be
    {H}:=\hbar\,\hat k\, \bone_3 \otimes\bsigma_3.
    \label{H-6-component}
    \ee
This is the operator acting on the elements $\left[\begin{array}{c}\bfxi \\ \bfzeta\end{array}\right]$ of ${\cH}$ according to ${H}\left[\begin{array}{c}\bfxi \\ \bfzeta\end{array}\right]=\hbar
\left[\begin{array}{c} \hat k\,\bfxi \\ -\hat k\,\bfzeta\end{array}\right]$.

Next, we introduce a pair of operators $U_{x_0^0},U'_{x_0^0}:\cV\to{\cH}$ that satisfy
    \begin{align}
    &U_{x_0^0}(\bfA):=\Psi(x_0^0), && U'_{x_0^0}:=\sqrt\ell \, D^{1/4}U_{x_0^0}=
    \sqrt\ell \,\hat k^{1/2}U_{x_0^0}.
    \label{U-zero-3}
    \end{align}
Then the generator $h$ of time translations together with the operators $H$, $U_{x_0^0}$, and $U'_{x_0^0}$ turn out to satisfy (\ref{h=def-2}) and (\ref{h-UHU-prime-2}). In particular, pulling back the inner product of ${\cH}$ via $U'_{x_0^0}$ produces an invariant positive-definite inner product $\cbr\cdot,\cdot\ckt$ that remains unchanged  under the Lorentz transformations mapping $\cV$ to $\cV$. Denoting the inner product of $\tilde\cH$ by $\bbr\cdot|\cdot\kkt$, we have the following analog of (\ref{rel-inn-prod-inv-2}).
    \bea
    \cbr\bfA,\tilde\bfA\ckt&:=&\bbr U'_{x_0^0}\bfA|U'_{x_0^0} \tilde\bfA\kkt
    =\frac{\ell }{2}\left[\bbr \bfA(x_0^0)|\hat k \tilde\bfA(x_0^0)\kkt+
    \bbr \dot\bfA(x_0^0)|\hat k^{-1}\dot{\tilde\bfA}(x_0^0)\kkt\right],
    \label{inner-product-3}
    \eea
We define the Hilbert space $\sH$ of the state vectors of the photon by giving this inner product to $\cV$. The $\ell$ appearing in (\ref{inner-product-3}) is a physically irrelevant free parameter of dimension $[{\rm Electric~Charge}/({\rm Length}\times{\rm Energy})]^2$ so that $\cbr\bfA,\tilde\bfA\ckt$ is dimensionless. Similarly to the parameter $\ell$ appearing in Section~\ref{Sec3}, $\ell$ drops out of the expression for the expectation value of observables. Therefore its value and even its dimension are physically irrelevant.

Unlike the case of scalar fields, Eq.~(\ref{inner-product-3}) does not imply that $U'_{x_0^0}:\sH\to{\cH}$ is a unitary operator. This is because the range of $U'_{x_0^0}$ consists only of those elements $\Phi$
%$\left[\begin{array}{c}\bfxi \\ \bfzeta\end{array}\right]$
of ${\cH}$ that fulfill the constraint $\hat\bfk\cdot\Phi=0$.
%$\hat\bfk\cdot\bfxi=\hat\bfk\cdot\bfzeta=0$.
Because these form a proper subset of ${\cH}$, $U'_{x_0^0}$ is not onto. We can still use $U'_{x_0^0}$ to determine the Hermitian operators acting in $\sH$ provided that we view $U'_{x_0^0}$ as an operator mapping $\sH$ onto its range. We denote this by $\cH'$; i.e., $\cH':={\rm Ran}(U'_{x_0^0})$.

\subsection{Hamiltonian, Helicity, Momentum, and Chirality Operators}
\label{Sec4B}

Consider the operator $\hat\fh:\tilde\cH\to\tilde\cH$ defined by
	\be
    \hat\fh:={\hat k}^{-1}\hat\bfk\cdot\bfS=
	i\,{\hat k}^{-1}
	\left[\begin{array}{ccc}
    0 & -\hat k_3 & \hat k_2\\
	\hat k_3 & 0 & -\hat k_1\\
	-\hat k_2 &\hat k_1 & 0
	\end{array}\right],
	\label{2-20}
	\ee
where the components of $\bfS$ are given by $S_1:=\lnd_7$, $S_2:=-\lnd_5$, and $S_3:=\lnd_2$, \cite{footnote4}. $\hat\fh$ is a Hermitian operator acting $\tilde\cH$ that satisfies
	\bea
	(\hat k\,\hat\fh\, \bfxi)(\bfx)&=&i(\hat{\bfk}\times\bfxi)(\bfx)=
	\bnabla\times\bfxi(\bfx),
	\label{curl=}\\
	\hat\fh^3&=&\hat\fh.
	\label{h3=h}
	\eea
As the latter equation suggests, the eigenvalues of  $\hat\fh$ are $0,\pm1$. Therefore we can obtain an orthogonal direct sum decomposition of $\tilde\cH$ into the eigenspaces of $\hat\fh$;
$\tilde\cH=\tilde\cH_0\oplus\tilde\cH_{-1}\oplus\tilde\cH_1$, where $\tilde\cH_{{s}}:=\{\bfxi\in\tilde\cH\:|\:\hat\fh\,\bfxi={s}\,\bfxi\}$ and ${s}=0,\pm1$. In particular, we can write every
$\bfzeta\in\tilde H$ in the form $\bfzeta=\bfzeta_0+\bfzeta_{-1}+\bfzeta_{1}$ for some $\bfzeta_{{s}}\in\tilde\cH_{{s}}$. Because $\hat\fh\,\bfzeta_{{s}}={s}\,\bfzeta_{{s}}$ and $\hat\bfk\cdot\hat\fh=0$, we have $\hat\bfk\cdot\bfzeta_\pm=0$. This in turn implies that $\hat\bfk\cdot\bfzeta=\hat\bfk\cdot\bfzeta_0$. Therefore the vectors $\bfzeta$ satisfying $\hat\bfk\cdot\bfzeta=0$ constitute $\tilde\cH_{-1}\oplus\tilde\cH_1$, and the range of $U'_{x_0^0}$ is given by
	\bea
	\cH'&=&\left\{\left.
	\left[\begin{array}{c}
	\bfxi\\ \bfzeta\end{array}\right]\in\cH~\right|~\bfxi,
    \bfzeta\in \tilde\cH_{-1}\oplus\tilde\cH_1\right\}
%    =(\tilde\cH_{-1}\oplus\tilde\cH_1)\otimes\C^2
    =(\tilde\cH_{-1}\otimes\C^2)\oplus(\tilde\cH_1\otimes\C^2)
    .
	\label{cH-prime=}
	\eea
The subspaces $\cH'_{\pm 1}:=\tilde\cH_{\pm1}\otimes\C^2$  appearing in this equation correspond to eigenspaces of the operator $\hat\fh\otimes\bone_2:\cH\to\cH$ with eigenvalue $\pm 1$. Clearly $\cH'$ is an invariant subspace of this operator \cite{footnote7}. Therefore we can restrict $\hat\fh\otimes\bone_2$ to $\cH'$ and view it as an operator acting in $\cH'$. Similarly, we treat $U_{x_0^0}^{\prime}$ as an operator mapping $\sH$ onto $\cH'$; $U_{x_0^0}^{\prime}:\sH\to\cH'$.
This makes it a unitary operator. Therefore we can use it to determine the Hermitian operators acting in $\sH$ from those acting in $\cH'$. The principal example is the generator of time translations,  $h:\sH\to\sH$. We can check that $\cH'_{\pm 1}$ are invariant subspaces of the operator $H$. Therefore, we can view $H$ as an operator acting in $\cH'$. Because this is a Hermitian operator that together with $h$ fulfill (\ref{h-UHU-prime-2}), $h$ is a Hermitian operator acting in $\sH$. We take it as the Hamiltonian of our quantum system; we use the pair $(\sH,h)$ to define the quantum system for a free photon.

Next, consider
	\be
	\Lfh:=U_{x_0^0}^{\prime -1}(\hat\fh\otimes{\bone_2}) \, U'_{x_0^0}=U_{x_0^0}^{-1}(\hat\fh
    \otimes{\bone_2}) U_{x_0^0}.
	\label{Hc-helicity}
	\ee
This is a Hermitian operator acting in $\sH$ which we identify with the helicity observable. It is easy to see that $\Lfh$ has two eigenvalues, namely $\pm 1$, and that the corresponding eigenspaces are given by $\sH_{\pm 1}:=U_{x_0^0}^{\prime -1}(\cH'_{\pm1})$. The state vectors of the photon residing in $\sH_{\pm 1}$ are said to have helicity $\pm 1$.

With the help of (\ref{h-UHU-prime-2}), (\ref{H-6-component}), (\ref{2-20}) and (\ref{Hc-helicity}), we can verify that
    \be
    [h,\Lfh]=0.
    \label{h-helicity-commute}
    \ee
This shows that there is a basis of $\sH$ consisting of common eigenvectors of $h$ and $\Lfh$. These correspond to the energy eigenstates of the photon that have a definite helicity. The energy and the helicity of such a state do not determine it in a unique manner. To specify such a state vector we need to use other observables that commute with $h$ and $\Lfh$.

According to (\ref{2-20}), $\hat\fh$ commutes with the operator $\hat k_j:\tilde\cH\to\tilde\cH$. This together with (\ref{Hc-helicity}) imply that $\Lfh$ commutes with the operators $K_j:\sH\to\sH$ given by
    \be
    K_j:=U_{x_0^0}^{\prime -1}(\hat k_j \otimes{\bone_2}) \, U'_{x_0^0}
    =U_{x_0^0}^{-1} (\hat k_j \otimes{\bone_2}) \, U_{x_0^0}.
	\label{momentum-3}
	\ee
Because $\hat k_j \otimes{\bone_2}$ is a Hermitian operator acting in $\cH'$ and $U'_{x_0^0}:\sH\to\cH'$ is a unitary operator, $K_j$ are Hermitian operators acting in $\sH$. We identify the components $P_j:\sH\to\sH$ of photon's momentum operator with $\hbar K_j$, i.e., take the momentum operator to be $\bfP:=\hbar\bK$, where $\bK:=(K_1,K_2,K_3)$.

It is not difficult to check that
    \be
    [K_j,\Lfh]=0.
    \label{momentum-helicity-commute}
    \ee
This relation implies the existence of a complete set of state vectors of the photon with definite momentum and helicity. These correspond to circularly polarized plane-wave solutions of the wave equation (\ref{dyn-eq-3}). We can express them as
    \be
    \bfA_{\bfk,\sigma}^\ep(x^0,\bfx)=N^\ep_{\bfk,\sigma}\,
	e^{-i\ep kx^0}\phi_\bfk(\bfx)\bfu_{\sigma}(\bfk),
    \label{plane-wave=1}
    \ee
where $\bfk$ and $\sigma$ respectively correspond to the eigenvalues of $\bK$ and $\Lfh$, $\epsilon$ gives the sign of energy, $N^\ep_{\bfk,\sigma}$ are normalization constants, $k:=|\bfk|$, $\phi_\bfk$ is defined by (\ref{phi=3}), and
    \bea
    \bfu_{\sigma}(\bfk)&:=&\f{1-\delta_{0 k_1}\delta_{0 k_2}}{k\sqrt{2(k_1^2+k_2^2)}}
    \left[\begin{array}{c}
    -k_1k_3+i\sigma k k_2 \\
    -k_2k_3-i\sigma k k_1 \\
    k_1^2+k_2^2
    \end{array}\right]+\frac{\delta_{0 k_1}\delta_{0 k_2}{\rm sgn}(k_3)}{\sqrt 2}
    \left[\begin{array}{c}
   1 \\
    i\sigma \\
    0
    \end{array}\right].
    \label{uk=3}
    \eea
Note that $\bfu_{\sigma}(\bfk)$ are the eigenvectors of the matrix
	\be
	\boldsymbol{\fh}:={k}^{-1}\bfk\cdot\bfS
	\label{cross}
	\ee
with eigenvalue $\sigma=\pm 1$. Together with $\bfu_0(\bfk):=\bfk/k$ they form an orthonormal basis of $\C^3$, i.e., for all $s,s'=0,\pm1$,
    \begin{align}
    &\bfu_{s}(\bfk)^\dagger\bfu_{s'}(\bfk)=\delta_{ss'},
    &&\sum_{\tilde s=-1}^1 \bfu_{\tilde s}(\bfk)\bfu_{\tilde s}(\bfk)^\dagger={\bone_3},
    \label{u-ortho}
    \end{align}
where we use $\dagger$ to denote the Hermitian conjugate (conjugate-transpose) of the corresponding column vector or matrix. We can also express (\ref{u-ortho}) in the form
    \begin{align}
    &  \bfu_{\sigma}(\bfk)^\dagger\bfu_{\sigma'}(\bfk)=\delta_{\sigma \sigma'},
    ~~~~~~~~~\bfk^\dagger\bfu_{\sigma}(\bfk)=\bfk\cdot \bfu_{\sigma}(\bfk)=0,
    \label{orthogonal-3}\\
    &  \sum_{\tilde\sigma=\pm1}u_{\tilde\sigma}(\bfk) u_{\tilde\sigma}(\bfk)^\dagger=
    {\bone_3}-k^{-2}\bfk\, \bfk^\dagger,
    \label{completeness-3}
    \end{align}
where $\sigma,\sigma'=\pm 1$ are arbitrary, and we treat $\bfk$ as a column vector.

Using (\ref{phi=3}), (\ref{inner-product-3}), (\ref{plane-wave=1}),  and (\ref{u-ortho}), we can show that
	$\cbr \bfA^{\epsilon}_{\bfk,\sigma}, \bfA^{\tilde\epsilon}_{\tilde\bfk,\tilde\sigma}\ckt=
	\ell \,k |N^\epsilon_{\bfk,\sigma}|^2 \,\delta_{\epsilon \tilde\epsilon}\,
	\delta_{\sigma \tilde\sigma}\,\delta^3(\bfk-\tilde{\bfk})$.
Therefore $\bfA^{\epsilon}_{\bfk,\sigma}$ form an orthonormal basis provided that $|N^\epsilon_{\bfk\sigma}|=1/\sqrt{\ell \, k}$. A convenient choice for the phase of $N^\epsilon_{\bfk\sigma}$ is $e^{i\epsilon k x_0^0}$. Making this choice, i.e., setting
	\be
	N^\epsilon_{\bfk\sigma}:=\frac{e^{i\epsilon k x_0^0}}{\sqrt{\ell \, k}},
	\label{N-eks=}
	\ee
and using (\ref{U-zero-3}), we find
    \be
    U'_{x_0^0}\bfA^\epsilon_{\bfk,\sigma}=\phi_{\bfk}\,\bfu_\sigma(\bfk)\otimes \bbe_\epsilon.
    \label{U-prime-basis-basis}
    \ee
This equation suggests that we identify $\epsilon$ with the eigenvalues of the chirality operator:
	\be
	C:=U_{x_0^0}^{\prime -1}(\bsigma_3\otimes{\bone_2}) U_{x_0^0}^{\prime}
	=U_{x_0^0}^{-1}(\bsigma_3\otimes{\bone_2})  U_{x_0^0}.
	\label{chirality-3}
	\ee
This is a Hermitian operator acting in $\sH$ which, in light of (\ref{A-c=def}), (\ref{6-comp-3}), and (\ref{U-zero-3}), admits the following explicit expression.
    \be
    C(\bfA)={\bfA}_c.
	\label{chirality=3}
    \ee
Furthermore, it satisfies
	\bea
    	&&C^2=1,~~~~~~~~~~h=C P,
	\label{C2=hCP}\\
    	&&\left[C,{\Lfh}\right]=\left[C,K_j\right]=\left[C, h\right]=0,
	\label{C-commutes}
	\eea
where $P:=\sqrt{\bfP^2}=\hbar\sqrt{\bK^2}$, and we have made use of (\ref{U-zero-3}), (\ref{h-UHU-prime-2}), (\ref{2-20}), (\ref{Hc-helicity}), (\ref{momentum-3}), and (\ref{chirality-3}).

As we mentioned earlier, $C$ and $\Lfh$ are commuting Hermitian operators acting in $\sH$. Therefore they have common eigenvectors $\bfA_\sigma^\epsilon$ satisfying
    \bea
    C\bfA_\sigma^\epsilon&=&\epsilon\,\bfA_\sigma^\epsilon,
    \label{C-eg-va-eq}\\
    \Lfh\bfA_\sigma^\epsilon&=&\sigma\,\bfA_\sigma^\epsilon.
    \label{Helicity-eg-va-eq}
    \eea
The following is a useful consequence of (\ref{U-zero-3}), (\ref{chirality=3}), and (\ref{C-eg-va-eq}).
    \be
    U'_{x_0^0}\bfA_\sigma^\epsilon=\sqrt{\ell }\,
    \hat k^{1/2}\bfA_\sigma^\epsilon(x_0^0)\otimes\bbe_\epsilon.
    \label{id-101}
    \ee
We can use it together with (\ref{maxwell-3}), (\ref{A-c=def}), (\ref{curl=}), (\ref{Hc-helicity}), (\ref{chirality-3}), (\ref{C-eg-va-eq}), (\ref{Helicity-eg-va-eq}), and (\ref{id-101}) to compute the electric and magnetic fields, $\bfE_\sigma^\epsilon(x^0):=-\dot\bfA_\sigma^\epsilon(x^0)$ and
$\bfB_\sigma^\epsilon(x^0):=\hat\bfk\times\bfA_\sigma^\epsilon(x^0)$, for $\bfA_\sigma^\epsilon$. In view of the fact that $x_0^0$ is an arbitrary real number, this yields
    \begin{align}
    & \bfE_\sigma^\epsilon(x^0)=i\epsilon\,\hat k\bfA_\sigma^\epsilon(x^0),
    && \bfB_\sigma^\epsilon(x^0)=\sigma\,\hat k\bfA_\sigma^\epsilon(x^0).
    \label{E-B=definite}
    \end{align}
As a consequence of these relations, we respectively find the following expressions for the energy density and Poynting vector \cite{Jackson} of a photon with a definite helicity and sign of energy.
    \bea
    u_\sigma^\epsilon(x^0,\bfx)&:=&
    \frac{1}{4\pi}\left\{\RE[\bfE_\sigma^\epsilon(x^0,\bfx)]^2+
    \RE[\bfB_\sigma^\epsilon(x^0,\bfx)]^2\right\}
    =\frac{1}{4\pi}\left|\bfE_\sigma^\epsilon(x^0,\bfx)\right|^2,
    \label{energy-density=}\\
    \boldsymbol{\cS}_\sigma^\epsilon(x^0,\bfx)&:=&
    \frac{c}{4\pi}\RE[\bfE_\sigma^\epsilon(x^0,\bfx)]\times
    \RE[\bfB_\sigma^\epsilon(x^0,\bfx)]=
    %\frac{\epsilon\,\sigma\,c}{4\pi}\,
    %\RE\left[\Big(\hat k  \bfA_\sigma^\epsilon(x^0) \Big)(\bfx)\right]\times
    %\IM\left[\Big(\hat k  \bfA_\sigma^\epsilon(x^0) \Big)(\bfx)\right]\nn\\
    %&=&
    \frac{i\epsilon\,\sigma\,c}{8\pi}\;\bfE_\sigma^\epsilon(x^0,\bfx)\times
    \bfE_\sigma^\epsilon(x^0,\bfx)^*.
    \label{poynting=}
    \eea

\subsection{Photon's position operator}
\label{Sec4C}

We begin our study of Photon's position operator $\bfX$ by demanding that its components $X_j$ act as Hermitian operators in the Hilbert space $\sH$ and fulfill the following two conditions.
	\begin{itemize}
	\item[ (i)] Together with the components $P_j$ of photon's momentum operator $\bfP$ they satisfy the canonical commutation relations (\ref{H-algebra}).
	\item[(ii)] They commute with both the helicity and chirality operators, i.e.,
	\be
	[X_j,\Lfh]=[X_j,C]=0.
	\label{X-h-C-commute}
	\ee
	\end{itemize}

In order to determine $X_j$, first we identify $\bfu_s(\bfk)$ as the columns of a $\bfk$-dependent matrix $\bU(\bfk)$, i.e.,
$\bU(\bfk):=\big[\:\bfu_{+1}(\bfk)\;~\bfu_{-1}(\bfk)\;~\bfu_0(\bfk)\:\big]$. Because $\bfu_s(\bfk)$ form an orthonormal basis of $\C^3$, $\bU(\bfk)$ is a unitary matrix. We use it to define a linear operator $\fU:\sH\to L^2(\R^3)\otimes\C^2\otimes\C^2$ according to
    \be
    \fU:=\bU(\hat\bfk)^\dagger U'_{x_0^0}=\sqrt{\ell }\,\hat k^{1/2}\bU(\hat\bfk)^\dagger U_{x_0^0}.
    \label{fU-prime}
    \ee
In view of (\ref{u-ortho}) and (\ref{U-prime-basis-basis}), it is easy to see that
    \be
	\fU\,\bfA^\epsilon_{\bfk,\sigma}=\phi_{\bfk}\,\bbe_\sigma\otimes \bbe_\epsilon.
	\label{fU-prime-basis}
	\ee
Because $\bfA^\epsilon_{\bfk,\sigma}$ and $\phi_{\bfk}\,\bbe_\sigma\otimes \bbe_\epsilon$ respectively form orthonormal bases of $\sH$ and $L^2(\R^3)\otimes\C^2\otimes\C^2$, this shows that $\fU$ is a unitary operator mapping $\sH$ to $L^2(\R^3)\otimes\C^2\otimes\C^2$. We use it to
define $X_j:\sH\to\sH$ according to
	\be
	X_j:=\fU^{-1}(\hat x_j {\bone_2}\otimes{\bone_2})\,\fU,
	\label{Xj-photon}
	\ee
where $\hat x_j$ are the components of the standard position operator acting in $L^2(\R^3)$.	
Clearly $\hat x_j {\bone_2}\otimes{\bone_2}$ is a Hermitian operator acting in $L^2(\R^3)\otimes\C^2\otimes C^2$. This together with the unitarity of $\fU$ and (\ref{Xj-photon}) establish the Hermiticty of $X_j$.

It is not difficult to show that we can also express $P_j$, $\Lfh$, and $C$ in the form
	\begin{align}
	&P_j=\fU^{-1}(\hat p_j {\bone_2}\otimes{\bone_2})\,\fU,
	&& \Lfh=\fU^{-1}(\bsigma_3\otimes{\bone_2})\,\fU,
	&&C=\fU^{-1}({\bone_2}\otimes\bsigma_3)\,\fU.
	\label{Pj-Fh-C-photon}
	\end{align}
Equations (\ref{Xj-photon}) and (\ref{Pj-Fh-C-photon}) ensure that $X_j$ satisfy Conditions (i) and (ii).
	
In order to obtain the explicit form of $X_j$ we compute its action on the basis vectors $\bfA_{\bfk,\sigma}^\epsilon$. In light of the fact that these form an orthonormal basis for $\sH$, we have
	\bea
	X_j\bfA_{\bfk,\sigma}^\epsilon&=&
	\sum_{\tilde\epsilon=\pm}\sum_{\tilde\sigma=\pm 1}\int_{\R^3}d^3\tilde\bfk\:
	\cbr\bfA_{\tilde\bfk,\tilde\sigma}^{\tilde\epsilon},X_j\bfA_{\bfk,\sigma}^\epsilon\ckt
	\bfA^{\tilde\epsilon}_{\tilde\bfk,\tilde\sigma}.
	\label{Xj=31}
	\eea
Because $\fU$ is a unitary operator,
	\bea
	\cbr\bfA_{\tilde\bfk,\tilde\sigma}^{\tilde\epsilon},X_j\bfA_{\bfk,\sigma}^\epsilon\ckt&=&
	\bbr\,\fU\,\bfA_{\tilde\bfk,\tilde\sigma}^{\tilde\epsilon}\,|
	\,\fU\,X_j\bfA_{\bfk,\sigma}^\epsilon\,\kkt
    %\nn\\&=&
    =\bbr\,\phi_{\tilde\bfk}\bbe_{\tilde\sigma}\otimes\bbe_{\tilde\epsilon}\,|\,\hat x_j\phi_{\bfk}\, 	 \bbe_\sigma\otimes\bbe_\epsilon\,
    \kkt\nn\\&=&
    -i\,\delta_{\tilde\sigma \sigma}\,\delta_{\tilde\epsilon \epsilon}\,
	\frac{\partial}{\partial k_j}\delta^3(\tilde\bfk-\bfk),
	\label{AXA-3}
	\eea
where we have employed (\ref{fU-prime-basis}) and (\ref{Xj-photon}). Substituting (\ref{AXA-3}) in (\ref{Xj=31}) and making use of (\ref{plane-wave=1}), (\ref{N-eks=}), (\ref{orthogonal-3}) and (\ref{completeness-3}), we obtain
	\[(X_j\bfA_{\bfk,\sigma}^\epsilon)(x^0,\bfx)=
	\left[\hat x_j+\frac{ik_j}{2k^2}-i(x^0-x_0^0)\frac{k_j}{k^2}\partial_0
	-i\left(\frac{\partial}{\partial k_j}\bfu_\sigma(\bfk)\right)\bfu_\sigma(\bfk)^\dagger\right]
	\bfA_{\bfk,\sigma}^\epsilon(x^0,\bfx).\]
This relation allows us to write the action of the position operator $\bfX:=(X_1,X_2,X_3)$ on the fields $\bfA(x^0,\bfx)$ as
    \be
    (\bfX\bfA)(x^0,\bfx)=\big(\boldsymbol{\fX}\bfA(x^0)\big)(\bfx),
    \label{position-33}
    \ee
where
	\be
	\boldsymbol{\fX}=
    \hat \bfx+\frac{i\hat \bfk}{2\hat k^2}-i(x^0-x_0^0)\frac{\hat \bfk}{\hat k^2}\,\partial_0
	-i\sum_{\sigma=\pm1}
	\left[\bnabla_{\hat\bfk}\bfu_\sigma(\hat\bfk)\right]\bfu_\sigma(\hat\bfk)^\dagger.
	\label{Photon-X=}
	\ee

Next, we compute the effect of the position operator $\bfX$ on the electric field $\bfE:=-\dot\bfA$. Let $\tilde\bfA_j:=X_j\bfA$. Then the electric field associated with $\tilde\bfA_j$ is given by $\tilde\bfE_j=-\dot{\tilde\bfA}_j$. Expressing $\tilde\bfE_j(x^0,\bfx)$ in the form $\big(\fX_j^{(E)}\bfE(x^0)\big)(\bfx)$ and using (\ref{position-33}) and (\ref{Photon-X=}) to compute $\boldsymbol{\fX}^{(E)}:=(\fX_1^{(E)},\fX_2^{(E)},\fX_3^{(E)})$, we find
    \be
    \boldsymbol{\fX}^{(E)}=\boldsymbol{\fX}-\frac{i\hat\bfk}{\hat k^2}=
     \hat \bfx-\frac{i\hat \bfk}{2\hat k^2}-i(x^0-x_0^0)\frac{\hat \bfk}{\hat k^2}\,\partial_0
	-i\sum_{\sigma=\pm1}
	\left[\bnabla_{\hat\bfk}\bfu_\sigma(\hat\bfk)\right]\bfu_\sigma(\hat\bfk)^\dagger.
	\label{Photon-X=E}
	\ee
With the help of (\ref{Photon-X=}), (\ref{position-33}), and (\ref{Photon-X=E}) we have verified that $\tilde\bfA_j$ satisfies the constraint (\ref{constract-A-dot}), the wave equation (\ref{dyn-eq-3}),  and the Coulomb gauge condition (\ref{TC-gauge}). This provides a highly nontrivial consistency check on our calculations.

We can also express $\boldsymbol{\fX}^{(E)}$ in terms of the linear polarization vectors:
$\bfa_1(\bfk):=\left[\bfu_{+1}(\bfk)+\bfu_{-1}(\bfk)\right]/\sqrt 2$ and $\bfa_2(\bfk):= -i \left[\bfu_{+1}(\bfk)-\bfu_{-1}(\bfk)\right]/\sqrt 2$.
%	\begin{align}
%	&\bfa_1(\bfk):=\frac{1}{\sqrt 2}\left[\bfu_{+1}(\bfk)+\bfu_{-1}(\bfk)\right],
%	&\bfa_2(\bfk):=\frac{-i}{\sqrt 2}\left[\bfu_{+1}(\bfk)-\bfu_{-1}(\bfk)\right].
%	\end{align}
These together with $\bfa_3(\bfk):=\bfu_0(\bfk)=\bfk/k$ form a complete orthonormal subset of $\C^3$ that satisfy
	\be
	\sum_{j=1}^3
	\left(\frac{\partial}{\partial k_j}\bfa_j(\bfk)\right)\bfa_j(\bfk)^\dagger
	=\sum_{s=-1}^1
	\left(\frac{\partial}{\partial k_j}\bfu_s(\bfk)\right)\bfu_s(\bfk)^\dagger.
	\label{u-a-3}
	\ee
Because $\hat\bfk\cdot \bfE(x^0)=0$, adding
the term $\left[\bnabla_{\hat\bfk}\bfu_0(\hat\bfk)\right]\bfu_0(\hat\bfk)^\dagger$ to the right hand side of (\ref{Photon-X=E}) does not change the action of $\boldsymbol{\fX}^{(E)}$ on $\bfE(x^0)$. With the help of this observation and Eqs.~(\ref{Photon-X=E}) and (\ref{u-a-3}), we have
	\be
	\boldsymbol{\fX}^{(E)}=\hat \bfx-\frac{i\hat \bfk}{2\hat k^2}
    -i(x^0-x_0^0)\frac{\hat \bfk}{\hat k^2}\,\partial_0
	-i\sum_{j=1}^3
	\left[\bnabla_{\hat\bfk}\bfa_j(\hat\bfk)\right]\bfa_j(\hat\bfk)^\dagger.
	\label{Photon-X=a}
	\ee	
For $x^0=x^0_0$, this formula reproduces the expression for Hawton's position operator \cite{11} that was originally obtained by adding an appropriate term to the Pryce's position operator \cite{24} to make its components commute. Here we obtain it following a systematic approach that involves constructing the Hilbert space $\sH$ of the state vectors of the photon, trying to obtain a unitary operator that maps $\sH$ onto the familiar Hilbert space $L^2(\R^3)\otimes\C^2\otimes\C^2$ (of a pair of distinguishable nonrelativistic spin 1/2 particles), and finally using this operator to pull back the usual position operator of nonrelativistic quantum mechanics to $\sH$.

Let us also determine the effect of our position operator $\bfX$ on the magnetic field $\bfB=\bnabla\times\bfA$. A similar analysis shows that if $\tilde\bfB_j$ is the magnetic field associated with $X_j\bfA$, we can express $\tilde\bfB_j(x^0,\bfx)$ in the form $\big(\fX_j^{B}\bfB(x^0))\big(\bfx)$ where $\fX_j^{B}$ are components of
    \be
    \boldsymbol{\fX}^{(B)}=\hat \bfx+\frac{3i\hat \bfk}{2\hat k^2}
    -i(x^0-x_0^0)\frac{\hat \bfk}{\hat k^2}\,\partial_0
	-i\sum_{\sigma=\pm1}
	\left[\bnabla_{\hat\bfk}\bfu_\sigma(\hat\bfk)\right]\bfu_\sigma(\hat\bfk)^\dagger.
    \nn
	\label{Photon-X=B}
	\ee

\subsection{Photon's localized states}
\label{Sec4D}

The localized state vectors determined by the position operator (\ref{position-33}) are given by
    \be
    \bbfA^\epsilon_{\bfy,\sigma}:=\fU^{-1}(\delta_{\bfy}\bbe_\sigma\otimes\bbe_\epsilon),
    \label{LS-def-3}
    \ee
where $\delta_{\bfy}$ is defined by (\ref{delta=3}). Because $\fU:\sH\to L^2(\R^3)\otimes\C^2\otimes\C^2$ is a unitary operator, $\bbfA^\epsilon_{\bfy\sigma}$ satisfy the following orthonormality and completeness relations
    \begin{align}
    &\cbr\bbfA^\epsilon_{\bfy ,\sigma}, \bbfA^{\tilde\epsilon}_{\tilde\bfy, \tilde\sigma}\ckt=
    \delta_{\epsilon \tilde\epsilon}\delta_{\sigma \tilde\sigma}\delta^3(\bfy-\tilde\bfy),\\
    &\sum_{\epsilon=\pm}\sum_{\sigma=\pm1}\int_{\R^3}d^3 \bfx\,
    \cbr\bbfA^{\epsilon}_{\bfx,\sigma}, \bfA \ckt\,\bbfA^{\epsilon}_{\bfx,\sigma}=\bfA,
    \label{x-expand}
    \end{align}
where $\bfA\in\sH$ is an arbitrary state vector.

In order to obtain the explicit form of the localized photon fields $\bbfA^\epsilon_{\bfy,\sigma}(x^0,\bfx)$, we expand them in the orthonormal basis consisting of  $\bfA^{\epsilon}_{\bfk,\sigma}$;
    \be
    \bbfA^\epsilon_{\bfy,\sigma}=\sum_{\tilde\epsilon=\pm}\sum_{\tilde\sigma=\pm1}
    \int_{\R^3}d^3\bfk\,
    \cbr \bfA^{\tilde\epsilon}_{\bfk,\tilde\sigma},\bbfA^\epsilon_{\bfy,\sigma}\ckt\,
    \bfA^{\tilde\epsilon}_{\bfk,\tilde\sigma}.
    \label{localized-301}
    \ee
In view of the unitarity of $\fU$,
    $\cbr \bfA^{\tilde\epsilon}_{\bfk,\tilde\sigma},\bbfA^\epsilon_{\bfy,\sigma}\ckt=
    \bbr\fU\,\bfA^{\tilde\epsilon}_{\bfk,\tilde\sigma}|\fU\,\bbfA^\epsilon_{\bfy,\sigma}\kkt=
    \delta_{\tilde\sigma \sigma}\,\delta_{\tilde\epsilon \epsilon}\,\br\phi_{\bfk}|\delta_{\bfy}\kt=
    \delta_{\tilde\sigma \sigma}\,\delta_{\tilde\epsilon \epsilon}\phi_{\bfk}(\bfy)^*$.
Inserting this equation in (\ref{localized-301}) and using (\ref{phi=3}), (\ref{plane-wave=1}),  and (\ref{N-eks=}), we find
    \bea
    \bbfA^\epsilon_{\bfy,\sigma}(x^0,\bfx)&=&
    \frac{1}{(2\pi)^3\sqrt\ell }\int_{\R^3}d^3\bfk\,
    k^{-1/2} e^{-i\epsilon k(x^0-x^0_0)} e^{i\bfk\cdot(\bfx-\bfy)}
    \bfu_\sigma(\bfk),
    \label{localized-302}
    \eea
where $\bfu_\sigma(\bfk)$ is given by (\ref{uk=3}). The evaluation of the integral on the right-hand side of (\ref{localized-302}) for $x^0\neq x^0_0$ turns out to be intractable. We leave the details of the calculation of this integral for $x^0=x^0_0$ to Appendix~A. Here we give its final result.

First, we note that the right-hand side of (\ref{localized-302}) is a function of $\bfr:=\bfx-\bfy$. We use spherical coordinates $(r,\theta,\varphi)$ to label $\bfr$, where $r$, $\theta$, and $\varphi$ are respectively the radial coordinate, polar angle, and azimuthal angle. Performing the integral in (\ref{localized-302}) we find
    \be
    \bbfA^\epsilon_{\bfy,\sigma}(x_0^0,\bfx)=\frac{1}{\sqrt\ell \,r^{5/2}}
    \left[\begin{array}{c}
    \cos\varphi\, T_1(\theta)+\sigma\,\sin\varphi\,T_2(\theta)\\
    \sin\varphi\, T_1(\theta)-\sigma\,\cos\varphi\,T_2(\theta)\\
    T_3(\theta)\end{array}\right],
    \label{LS-401}
    \ee
where
    \bea
    T_1(\theta)&:=&\frac{5\Gamma(\frac{5}{4})^2}{8\sqrt 2\pi^2}\sin(2\theta)\Big[
    2\, _2 F_1(\mbox{$\frac{1}{4},\frac{1}{2};1;\sin^2\theta$})-
    \cos^2\theta\, _2F_1(\mbox{$\frac{3}{2},\frac{9}{4};2;\sin^2\theta$})+
    \sin^2\theta\, _2F_1(\mbox{$\frac{5}{2},\frac{3}{2};2;\sin^2\theta$})\Big],
    \nn%\label{T1=}
    \\
    T_2(\theta)&:=&\frac{3\Gamma(\frac{3}{4})^2}{16\sqrt 2\pi^2}\sin\theta \Big[
    -2\, _2 F_1(\mbox{$\frac{1}{2},\frac{3}{4};1;\sin^2\theta$})+
    \cos^2\theta\, _2F_1(\mbox{$\frac{3}{2},\frac{7}{4};2;\sin^2\theta$})\Big],
    \nn%\label{T2=}
    \\
    T_3(\theta)&:=&\frac{\Gamma(\frac{1}{4})^2}{64\pi\Gamma(\frac{3}{4})} \Big[
    [3+5\cos(2\theta)]\, _2 F_1(\mbox{$\frac{1}{4},\frac{1}{2};1;\sin^2\theta$})+
    \cos^2\theta[1-5\cos(2\theta)]\, _2F_1(\mbox{$\frac{5}{2},\frac{3}{2};2;\sin^2\theta$})\nn\\
    &&\hspace{1.5cm} -\mbox{$\frac{15}{4}$}\, _2F_1(\mbox{$\frac{9}{2},\frac{5}{2};3;\sin^2\theta$})\Big],
    \nn%\label{T3=}
    \eea
and $\Gamma$ and $_2 F_1$ stand for the Euler's Gamma function and Gauss's Hypergeometric function \cite{GR}, respectively. An unexpected outcome of the above formulas is that $T_1(\theta)$ and $T_2(\theta)$ diverge at $\theta=\pi/2$ while $T_3(\theta)$ is continuous but non-differentiable at this point. This shows that $\bbfA^\epsilon_{\bfy,\sigma}(x_0^0,\bfx)$ has a singularity not only at the point $r=0$ but on the whole equatorial plane $\theta=\pi/2$. This is in sharp contrast to the localized states of massive and massless scalar fields (of Sec.~\ref{Sec3}) and the Proca field \cite{51} which only diverge at $r=0$. Another notable observation is that the initial value of the localized state vectors for the photon (\ref{LS-401}) has the same $r$-dependence as those of a massless scalar field (\ref{LS-KG-zero-zero}); both are proportional to $r^{-5/2}$. This agrees with the asymptotic ($r\to\infty$) behavior of the weakly localized states of Refs.~\cite{34,35,pike}. Note however that (\ref{LS-401}) is an exact expression valid for large as well as small values of $r$. Furthermore, in contrast to the weakly localized states of \cite{34,35,pike}, our localized states describe a photon with a definite helicity that is localized at a single point in space.

Because the localized states given by (\ref{LS-401}) have definite helicity and sign of energy, we can use (\ref{E-B=definite}), (\ref{energy-density=}), and (\ref{poynting=}) to express the corresponding electric field, magnetic field, energy density, and Poynting vector as
    \bea
    &&\boldsymbol{\cE}_{\bfy,\sigma}^\epsilon(x^0,\bfx)=i\epsilon\big(\hat k\bbfA_{\bfy,\sigma}^\epsilon(x^0)\big)(\bfx)=
    \frac{i\epsilon}{(2\pi)^3\sqrt\ell }\int_{\R^3}d^3\bfk\,
    k^{1/2} e^{-i\epsilon k(x^0-x^0_0)} e^{i\bfk\cdot(\bfx-\bfy)}
    \bfu_\sigma(\bfk),
    \label{localized-E=}\\
    &&\boldsymbol{\cB}_{\bfy,\sigma}^\epsilon(x^0,\bfx)=\sigma\big(\hat k\bbfA_{\bfy,\sigma}^\epsilon(x^0)\big)(\bfx)
    =-i\epsilon\,\sigma\,\boldsymbol{\cE}_{\bfy,\sigma}^\epsilon(x^0,\bfx),
    \label{localized-B=}\\
    &&u_{\bfy,\sigma}^\epsilon(x^0,\bfx)=
    \frac{1}{4\pi}\,\left|\boldsymbol{\cE}_{\bfy,\sigma}^\epsilon(x^0,\bfx)
    \right|^2,
    \label{localized-u=}\\
    &&\boldsymbol{\cS}_{\bfy,\sigma}^\epsilon(x^0,\bfx)=\frac{i\epsilon\,\sigma\,c}{8\pi}\,
    \boldsymbol{\cE}_{\bfy,\sigma}^\epsilon(x^0,\bfx)\times \boldsymbol{\cE}_{\bfy,\sigma}^\epsilon(x^0,\bfx)^*.
    \label{localized-S=}
    \eea
respectively. Again, we were unable to obtain an analytic expression for the integral in (\ref{localized-E=}) except for $x^0=x^0_0$. For the latter case, we have
    \be
    \boldsymbol{\cE}_{\bfy,\sigma}^\epsilon(x^0_0,\bfx)=\frac{i\epsilon}{\sqrt\ell \,r^{7/2}}
    \left[\begin{array}{c}
    \cos\varphi\, T_4(\theta)+\sigma\,\sin\varphi\,T_5(\theta)\\
    \sin\varphi\, T_4(\theta)-\sigma\,\cos\varphi\,T_5(\theta)\\
    T_6(\theta)\end{array}\right],
    \label{LS-401-E}
    \ee
where
    \bea
    T_4(\theta)&:=&\frac{21 \Gamma(\frac{3}{4})^2}{1024\sqrt 2\pi^2}\sin(2\theta)\Big[
    32\, _2 F_1(\mbox{$\frac{1}{2},\frac{3}{4};1;\sin^2\theta$})-
    16\cos(2\theta)\, _2F_1(\mbox{$\frac{3}{2},\frac{7}{4};2;\sin^2\theta$})
    \nn\\
    &&\hspace{2.8cm}-3\sin^2(2\theta)\, _2F_1(\mbox{$\frac{5}{2},\frac{11}{4};3;\sin^2\theta$})\Big],
    \nn%\label{T1=}
    \\
    T_5(\theta)&:=&\frac{-5\Gamma(\frac{1}{4})^2}{1024\sqrt 2\pi^2}\sin\theta \Big[
    16\, _2 F_1(\mbox{$\frac{1}{4},\frac{1}{2};1;\sin^2\theta$})
    -4[7+11\cos(2\theta)]\, _2F_1(\mbox{$\frac{5}{4},\frac{3}{2};2;\sin^2\theta$})\nn\\
    &&\hspace{2.3cm}
    -3\cos^2\theta[3-19\cos(2\theta)]\, _2 F_1(\mbox{$\frac{9}{4},\frac{5}{2};3;\sin^2\theta$})
    +45\cos^4\theta\sin^2\theta\, _2 F_1(\mbox{$\frac{13}{4},\frac{7}{2};4;\sin^2\theta$})\Big],
    \nn%\label{T2=}
    \\
    T_6(\theta)&:=&\frac{3\Gamma(\frac{3}{4})}{128\pi\Gamma(\frac{1}{4})} \Big[
    4[1+7\cos(2\theta)]\, _2 F_1(\mbox{$\frac{1}{2},\frac{3}{4};1;\sin^2\theta$})+
    4\cos^2\theta[3-7\cos(2\theta)]\, _2F_1(\mbox{$\frac{3}{2},\frac{7}{4};2;\sin^2\theta$})\nn\\
    &&\hspace{1.7cm}
    -21\cos^4\theta\sin^2\theta\, _2F_1(\mbox{$\frac{5}{2},\frac{11}{4};3;\sin^2\theta$})\Big].
    \nn%\label{T3=}
    \eea

\begin{figure}
	\begin{center}
	\includegraphics[scale=.6]{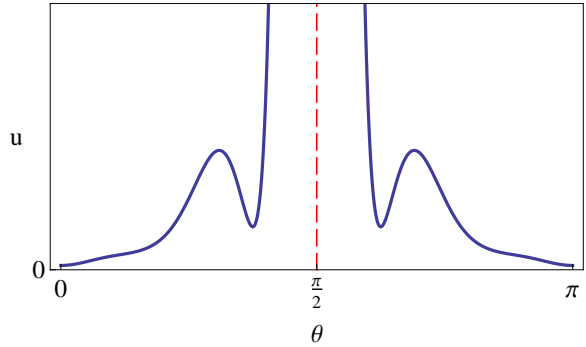}
	\caption{Graph of the energy density of a localized photon with definite helicity and sign of energy as $u:=u_{\bfy,\sigma}^\epsilon(x_0^0,\bfx)$  as a function of the polar angle $\theta$ for any nonzero value of the radial coordinate $r=|\bfx-\bfy|$. The dashed red line marks the plane $\theta=\pi/2$ on which $u$ blows up.}
	\label{fig1}
	\end{center}
	\end{figure}

According to (\ref{localized-B=}) -- (\ref{LS-401-E}),
    \begin{align}
    &u_{\bfy,\sigma}^\epsilon(x_0^0,\bfx)=\frac{1}{4\pi\,\ell \,r^7}\sum_{j=4}^6T_j(\theta)^2,
    &&\boldsymbol{\cS}_{\bfy,\sigma}^\epsilon(x_0^0,\bfx)=\mathbf{0}.
    \label{localized-u-S}
    \end{align}
We also observe that, for $x^0=x^0_0$, $\boldsymbol{\cE}_{\bfy,\sigma}^\epsilon$, $\boldsymbol{\cB}_{\bfy,\sigma}^\epsilon$, and $u_{\bfy,\sigma}^\epsilon$ blow up on the plane $\theta=\pi/2$, and that $u_{\bfy,\sigma}^\epsilon$ is independent of $\epsilon$, $\sigma$, and $\varphi$. In particular, it is axially symmetric with respect to the $x_3$-axis. Figure~\ref{fig1} shows the plot of $u_{\bfy,\sigma}^\epsilon(x_0^0,\bfx)$ as a function of $\theta$ (for fixed nonzero values of $r$).
	
The fact that the localized photon fields with a definite sign of energy and helicity turn out to be singular on a plane rather than a single point is one of the main results of the present investigation. It simply means that such a localized photon specifies a preferred axis (straight line) in space, namely the $x_3$-axis in the coordinate system we have adopted. To achieve a conceptual understanding of this result, we note that according to  (\ref{LS-401-E}) we can express the electric field $\boldsymbol{\cE}_{\bfy,-1}^\epsilon(x^0_0,\bfx)$ by rotating its value for $\varphi=0$ by the angle $\varphi$ about the $x_3$-axis (in the counterclockwise sense). But to get $\boldsymbol{\cE}_{\bfy,+1}^\epsilon(x^0_0,\bfx)$ from its value for $\varphi=0$, we need to affect a parity transformation that flips the sign of the $y$-component (equivalently the direction of the $y$-axis) and then perform the same rotation. This provides a geometric interpretation of having positive or negative helicity provided that we view the $x_3$-axis as a preferred axis with the following basic properties.
	\begin{itemize}
	\item[]{\rm $R_1$}) It passes through the center of the localized field, i.e., $\bfx=\bfy$, and
	\item[]{\rm $R_2$})  It is normal to the plane at which this field blows up.
	\end{itemize}
We term this axis the ``symmetry axis'' of the localized states, because their energy density is axially symmetric with respect to this axis.

The coincidence of the symmetry axis of our localized states with the $x_3$-axis has to do with our  use the formula (\ref{uk=3}) for the polarization vectors $\bfu_\sigma(\bfk)$. The fact that these are the eigenvectors of $\boldsymbol{\fh}:={k}^{-1}\bfk\cdot\bfS$ and that they fulfill the orthonormality and completeness relations (\ref{orthogonal-3}) and (\ref{completeness-3}) do not determine them uniquely; we can always multiply them by phase factors. The choice of these phase factors, which may in general depend on $\bfk$, affects the outcome of the calculation of the vector potential and electromagnetic field of the localized state, and in particular the plane at which they diverge.

In order to clarify the situation, we first recall the following more general form of $\bfu_\sigma(\bfk)$ whose derivation we give in Appendix~B.
    	\be
    	\bfu_\sigma(\bfk)=\frac{k^2\bfm-(\bfk\cdot\bfm)\bfk+i\sigma\,k(\bfk\times
	\bfm)}{k\sqrt{2[k^2-(\bfk\cdot\bfm)^2]}},
	\label{bfuk=gen}
	\ee
where $\bfm$ is an arbitrary unit vector not parallel to $\bfk$, \cite{footnote101}. If we set $\bfm=\bbe_{0}$ (i.e., the unit vector along the $x_3$-axis), Eq.~(\ref{bfuk=gen}) reduces to (\ref{uk=3}) for $k_1^2+k_2^2\neq 0$. If we do not make a particular choice for $\bfm$,
%\cite{footnote9}
the calculation of $\bbfA^\epsilon_{\bfy,\sigma}(x_0^0,\bfx)$ and $\boldsymbol{\cE}_{\bfy,\sigma}^\epsilon(x^0_0,\bfx)$ becomes quite complicated. We can nevertheless infer the singularity structure of the outcome, because we can always choose the coordinate system in which we perform this calculation such that its $x_3$-axis is along $\bfm$. In this coordinate system, Eqs.~(\ref{LS-401}) and (\ref{LS-401-E}) hold and $\bbfA^\epsilon_{\bfy,\sigma}(x_0^0,\bfx)$ and $\boldsymbol{\cE}_{\bfy,\sigma}^\epsilon(x^0_0,\bfx)$ diverge in the $x_1$-$x_2$ plane, i.e., the plane normal to $\bfm$. This argument shows that the symmetry axis fulfilling the requirements $R_1$ and $R_2$ is parallel to $\bfm$.

Next, let us denote the right-hand side of  (\ref{bfuk=gen}) by $\bfu_\sigma^{\bfm}(\bfk)$, so that $\bfu_\sigma^{\bbe_0}(\bfk)$ is given by (\ref{uk=3}). By construction $\bfu_\sigma^{\bfm}(\bfk)$ and $\bfu_\sigma^{\bbe_0}(\bfk)$ are normalized eigenvectors of $\boldsymbol{\fh}$ with eigenvalue $\sigma$. Because eigenvalues of $\boldsymbol{\fh}$ are nondegenerate, $\bfu_\sigma^{\bfm}(\bfk)$ and $\bfu_\sigma^{\bbe_0}(\bfk)$ can differ only by a possibly $\bfk$-dependent phase factor;
	\be
	\bfu_\sigma^{\bfm}(\bfk)=e^{i\phi_\sigma^{\bfm}(\bfk)}\bfu_\sigma^{\bbe_0}(\bfk).
	\label{gauge-01}
	\ee
In Appendix~B we show that
    \be
	e^{i\phi_\sigma^\bfm(\bfk)}=\frac{k^2m_3-k_3\bfk\cdot\bfm+i\sigma k(k_1m_2-k_2m_1)}{
	\sqrt{(k_1^2+k_2^2)[k^2-(\bfk\cdot\bfm)^2]}},
	\label{phase=}
	\ee
where $m_j$ are the components of $\bfm$.

Next, we explore the consequences of (\ref{gauge-01}) for the determination of the position operator (\ref{position-33}). To do this first we introduce the symbol $\bfX^{\bfm}$ to label the position operator given by (\ref{position-33}) and (\ref{Photon-X=}) with $\bfu_\sigma(\bfk)=\bfu_\sigma^{\bfm}(\bfk)$. Then in view of (\ref{Photon-X=}) and (\ref{gauge-01}), we have
	\be
	\bfX=\bfX^{\bfm}=\bfX^{\bbe_0}+\boldsymbol{\Theta}^{\bfm},
	\label{gauge-02}
	\ee
where
	\[\big(\boldsymbol{\Theta}^{\bfm}\bfA\big)(x^0,\bfx)
	:=\sum_{\sigma=\pm1}\big((\bnabla_{\hat\bfk}\phi_\sigma^\bfm(\hat\bfk))
	\bfu^{\bbe_0}_\sigma(\hat\bfk)\bfu^{\bbe_0}_\sigma(\hat\bfk)^\dagger\bfA(x^0)\big)(\bfx).\]
Equation (\ref{gauge-02}) reveals the $\bfm$-dependence of the photon position operator (\ref{position-33}). It shows that we have indeed an infinite family of admissible position operators parameterized by the unit vectors $\bfm$. This explains the origin of the choice of the symmetry axis of the localized states.

The existence of an infinite family of photon position operators might sound problematic, but it is really not very different from the existence of the family of position operators $\bfX_{\bfa}:=\bfX-\bfa 1$, where $\bfa\in\R^3$ and $1$ is the identity operator. Components of $\bfX^{\bfm}$ and $\bfP$ provide unitary-equivalent irreducible representations of the Heisenberg algebra for all $\bfm$. It is a choice of the observer to use any one of these equivalent representations. The fact that the symmetry axis of the localized states depends on the choice of this representation does not have any physical consequences, for the localized states are not observable quantities.

According to the above analysis, the photon position operator is determined by a unit vector $\bfm$ which in turn identifies the symmetry axis of the corresponding localized state vectors $\bbfA^\epsilon_{\bfx,\sigma}$. The latter form a position basis $\{\bbfA^\epsilon_{\bfx,\sigma}\}$ that we can employ for developing a position representation of the quantum mechanics of the photon.

\subsection{Photon's position wave function and probability density}
\label{Sec4E}

Following our discussion of the position wave function for the scalar fields, we identify the position wave function for a photon field $\bfA\in\sH$ with the coefficient,
	\be
	f(\epsilon,\sigma,\bfx):=\cbr \bbfA^\epsilon_{\bfx,\sigma},\bfA\ckt,
	\label{x-wf=def}
	\ee
of the expansion (\ref{x-expand}) of $\bfA$ in the (position) basis consisting of the localized state vectors $\bbfA^\epsilon_{\bfx,\sigma}$. Before exploring the properties of photon's position wave function, we wish to stress that because the localized fields $\bbfA^\epsilon_{\bfx,\sigma}$ have an implicit dependence on the choice of a symmetry axis, the same applies for the position wave function (\ref{x-wf=def}). In what follows we identify the symmetry axis with the $x_3$-axis, so that the expression (\ref{uk=3}) for the polarization vectors $\bfu_\sigma(\bfk)$ holds.

In order to obtain an explicit formula for $f(\epsilon,\sigma,\bfx)$, we first note that
	\be
	\bfu_\sigma(\hat\bfk)^\dagger\bfB(x_0^0)=
	\bfu_\sigma(\hat\bfk)^\dagger[\hat\bfk\times\bfA(x_0^0)]=
	\hat k\,\bfu_\sigma(\hat\bfk)^\dagger\hat{\fh}\bfA(x_0^0)=
	\sigma\,\hat k\,\bfu_\sigma(\hat\bfk)^\dagger\bfA(x_0^0),
	\label{uB=uA}
	\ee
where we have used (\ref{curl=}) and the fact that $\hat\fh \bfu_\sigma(\hat\bfk)=\sigma
\bfu_\sigma(\hat\bfk)$. Because $\fU:\sH\to L^2(\R^3)\otimes\C^2\otimes\C^2$ is a unitary operator, Eqs.~(\ref{LS-def-3}) and (\ref{uB=uA}) imply
	\bea
	f(\epsilon,\sigma,\bfx)&=&-\frac{i\sqrt\ell}{2}\,\Big(\hat k^{-1/2}\bfu_\sigma(\hat \bfk)^\dagger
    \left[\epsilon\,\bfE(x_0^0)+i\,\sigma\,\bfB(x_0^0)\right]\Big)(\bfx)
    \label{f=QRS-1}\\
    &=&\int_{\R^3} d\bfx^{\prime 3}\bfQ_\sigma(\bfx-\bfx')
    \left[\epsilon\,\bfE(x_0^0,\bfx')+i\,\sigma\,\bfB(x_0^0,\bfx')\right],
	\label{f=QRS}
	\eea
where $\bfQ_\sigma(\bfx-\bfx'):=
    - [i/2(2\pi)^3\sqrt\ell]\int_{\R^3} d\bfk^3 k^{-1/2}e^{i\bfk\cdot(\bfx-\bfx')}\bfu_\sigma(\bfk)^\dagger$.
We can write the integral on the right-hand side of this relation as the transpose of $\mathbf{I}^-_{-\sigma}(\bfx-\bfx')$ where $\mathbf{I}^-_{\sigma}(\bfr)$ is defined and evaluated in Appendix~A. See~Eq.~(\ref{Ipm=}) below.

Equation~(\ref{f=QRS}) relates the position wave function $f(\epsilon,\sigma,\bfx)$ to the Riemann-Silberstein wave function:
    \be
    \boldsymbol{\psi}_{RS}(\epsilon,\sigma,\bfx):=\epsilon\,\bfE(x_0^0,\bfx)+i\,\sigma\,\bfB(x_0^0,\bfx).
    \label{RS=exp}
    \ee
We can also relate $f(\epsilon,\sigma,\bfx)$ to the Landau-Peierls wave function $\boldsymbol{\psi}_{LP}$ which satisfies
    \be
    \boldsymbol{\psi}_{LP}(\epsilon,\sigma,\bfx)=(\hat k^{-1/2}\boldsymbol{\psi}_{RS})(\epsilon,\sigma,\bfx)=
    \pi\int_{\R^3} d^3\bfx'\,\frac{\boldsymbol{\psi}_{RS}(\epsilon,\sigma,\bfx')}{(2\pi|\bfx'-\bfx|)^{5/2}}.
    \label{LP=exp}
    \ee
In view of (\ref{f=QRS}) -- (\ref{LP=exp}), we have
    \be
    f(\epsilon,\sigma,\bfx)=-\frac{i\sqrt\ell}{2}\,\big(\bfu_\sigma(\hat \bfk)^\dagger
    \boldsymbol{\psi}_{LP}\big)(\epsilon,\sigma,\bfx).
    \label{f=Psi-LP}
    \ee

Having determined the explicit form of photon's position wave functions we can pursue the approach of Subsec.~\ref{Sec3C} to formulate a position representation for its quantum mechanics. The first step in this direction is the observation that for a given pair of photon fields $\bfA,\tilde\bfA\in\sH$ with position wave functions $f,\tilde f$, we have
    \[\cbr\bfA,\tilde\bfA\ckt=\sum_{\epsilon=\pm}\sum_{\sigma=\pm1}\int_{\R^3}d\bfx^3
    f(\epsilon,\sigma,\bfx)^*\tilde f(\epsilon,\sigma,\bfx)=\sum_{\epsilon=\pm}\sum_{\sigma=\pm1}
    \br f(\epsilon,\sigma,\cdot)|\tilde f(\epsilon,\sigma,\cdot)\kt.\]
In light of this relation we can view $f$ as an element of $L^2(\R^3)\otimes\C^2\otimes\C^2$ and identify the observables of the photon with Hermitian operators acting in this Hilbert space. This determines the position representation of the quantum mechanics of the photon, where the chirality, helicity, position, and momentum operators act on position wave functions $f(\epsilon,\sigma,\bfx)$ to give $\epsilon f(\epsilon,\sigma,\bfx)$, $\sigma f(\epsilon,\sigma,\bfx)$,
$\bfx f(\epsilon,\sigma,\bfx)$, and $-i\hbar\boldsymbol{\nabla}f(\epsilon,\sigma,\bfx)$, respectively. An important feature of this representation is its gauge-invariance. This is a straightforward consequence of Eq.~(\ref{f=QRS}) which shows the gauge-invariance of the position wave functions $f(\epsilon,\sigma,\bfx)$.

In the position representation, a localized state of the photon with center $\bfx'$, sign of energy $\epsilon'$, and helicity $\sigma'$ is described by the position wave function $f(\epsilon,\sigma,\bfx)=\delta_{\epsilon\epsilon'}\delta_{\sigma\sigma'}\delta^3(\bfx-\bfx')$.

Next, we examine the position wave function for a field with definite momentum. Let $\bfA_\kappa\in\sH$ correspond to the plane wave determined by the initial electromagnetic field
    \begin{align}
    &\bfE(x^0_0,\bfx)=E_0 e^{i {\kappa} x_3} \bbe_{1},
    &&\bfB(x^0_0,\bfx)=B_0 e^{i {\kappa} x_3} \bbe_{-1},
    \label{plane-wave=12}
    \end{align}
where $E_0$ and $B_0$ are complex coefficients, ${\kappa}$ is a positive real (wave)number, and $\bbe_s$ are defined in (\ref{0-4}). Substituting (\ref{plane-wave=12}) in (\ref{RS=exp}) -- (\ref{f=Psi-LP}) and making use of Eq.~(\ref{uk=3}) and the fact that the action of $\hat\bfk$ on $e^{i {\kappa} x_3}$ gives ${\kappa} e^{i{\kappa}x_3}\bbe_{0}$, we have $\boldsymbol{\psi}_{RS}(\epsilon,\sigma,\bfx)=e^{i{\kappa}x_3}\left(\epsilon E_0\,\bbe_1+i\sigma B_0\,\bbe_{-1}\right)$, $\boldsymbol{\psi}_{LP}(\epsilon,\sigma,\bfx)=
e^{i{\kappa}x_3} \left(\epsilon E_0\,\bbe_1+i\sigma B_0\,\bbe_{-1}\right)/\sqrt\kappa$,
and $f(\epsilon,\sigma,\bfx)=-i\sqrt\ell(\epsilon E_0+B_0)e^{i{\kappa} x_3}/2\sqrt{2{\kappa}}$. The following are simple consequences of the latter equation.
    \begin{enumerate}
    \item $-i\hbar\boldsymbol{\nabla}f(\epsilon,\sigma,\bfx)=\hbar\kappa
f(\epsilon,\sigma,\bfx)\bbe_0$. This shows that this field has a definite momentum $(0,0,\hbar {\kappa})$.
    \item $f(\epsilon,-1,\bfx)=f(\epsilon,+1,\bfx)$. This shows that the field does not have a definite helicity, which agrees with the fact that it is linearly polarized.
    \item For $E_0=\pm B_0$, $f(\mp,\sigma,\bfx)=0$. This shows that in this case the field has a definite sign of energy, namely $\pm$.
    \end{enumerate}

We conclude this section by a discussion of the probability density of spatial localization of a photon field $\bfA\in\sH$. In terms of the position wave function $f(\epsilon,\sigma,\bfx)$ of $\bfA$, this is given by
    \be
    \rho(\bfx)=\sum_{\epsilon=\pm}\sum_{\sigma=\pm1}\rho_{\epsilon,\sigma}(\bfx),
    \ee
where
    \be
    \rho_{\epsilon,\sigma}(\bfx):=
    \frac{|\cbr\bbfA^\epsilon_{\bfx,\sigma},\bfA\ckt|^2}{\cbr\bfA,\bfA\ckt}=
    \frac{|f(\epsilon,\sigma,\bfx)|^2}{\sum_{\epsilon=\pm}\sum_{\sigma=\pm1}
    \br f(\epsilon,\sigma,\cdot)|f(\epsilon,\sigma,\cdot)\kt}.
    \ee
This quantity is the probability density of the localization of a photon with sign of energy $\epsilon$ and helicity $\sigma$. We can use it to identify the probability of finding the values $\epsilon$ and $\sigma$ for a measurement of the chirality and helicity of the field $\bfA$ with $\sum_{\sigma=\pm1}\int_{\R^3}d^3\bfx\, \rho_{\epsilon,\sigma}(\bfx)$ and $\sum_{\epsilon=\pm} \int_{\R^3}d^3\bfx\, \rho_{\epsilon,\sigma}(\bfx)$, respectively.

Notice that because the position wave function $f(\epsilon,\sigma,\bfx)$ is uniquely determined by the electric and magnetic fields associated with $\bfA$, the above probability densities and probabilities are gauge-invariant quantities. Furthermore, we can follow the approach of \cite{51} to relate the conservation of the total probability to the existence of a locally conserved complex $4$-vector current density.

\begin{figure}
	\begin{center}
	\includegraphics[scale=.8]{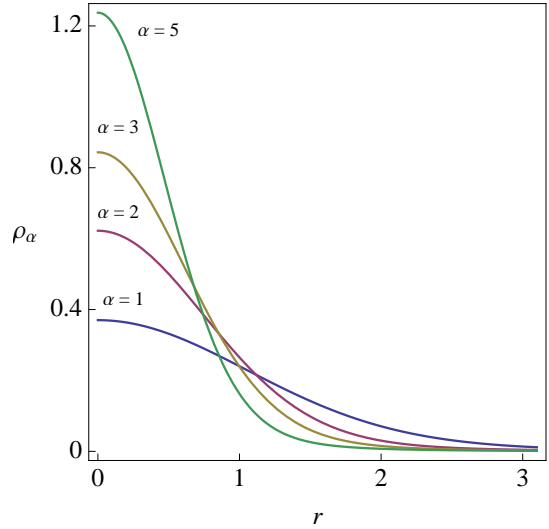}
	\caption{Graph of $\rho_\alpha$ as a function of $r$ for $\alpha=1,2,3$, and $5$. The values of $r$ and $\alpha$ are respectively given in units of $d$ and $d^{-2}$ where $d$ is an arbitrary unit of length.}
	\label{fig2}
	\end{center}
	\end{figure}

As a simple example, consider the Gaussian profile:
	\begin{align}
	&\bfA(x_0^0,\bfx)=0,
	&&\bfE(x_0^0,\bfx)=-\dot\bfA(x_0^0,\bfx)=E_0 e^{-\alpha r^2/2}(-x_2 \bbe_1+x_1\bbe_{-1}),
	\label{AE-zero}
	\end{align}
where $\alpha$ is a positive real constant, $E_0$ is a real or complex constant, and $r=|\bfx|$. Because
	\be
	\bfE(x_0^0,\bfx)=\boldsymbol{\nabla}\times\bfV(\bfx)~~~{\rm for}~~~\bfV(\bfx):=\alpha^{-1}E_0\, e^{-\alpha r^2/2}\bbe_0,
	\label{E-zero=}
	\ee
we have $\boldsymbol{\nabla}\cdot \bfE(x_0^0,\bfx)=0$. Therefore,
(\ref{AE-zero}) determines a state vector of the photon, $\bfA\in\sH$. In Appendix~C, we compute the position wave function associated with this state vector. The result is
	\be
	f(\epsilon,\sigma,\bfx)=\frac{\sqrt{\pi\ell}\,\Gamma(\frac{7}{4})\,\epsilon\,\sigma\,E_0}{
	64\;2^{\frac{1}{4}}\alpha^{\frac{3}{4}}}
	\left[16\, L_{-\frac{7}{4}}(-\mbox{$\frac{\alpha r^2}{2}$})+
	7\,\alpha\, r^2\: _1\!F_1(\mbox{$\frac{11}{4}$},3,\mbox{$-\frac{\alpha r^2}{2}$})\right],
	\label{fw=eg}
	\ee
where $L_\nu$ and $\: _1\!F_1$ are respectively the Laguerre function and Kummer confluent hypergeometric function \cite{GR}. In view of (\ref{fw=eg}), $\rho_{\epsilon,\sigma}(\bfx)=\epsilon\,\sigma\,\rho_\alpha(r)$ where $\rho_\alpha$ is an entire function of $r$. Figure~\ref{fig2} shows the graph of $\rho_\alpha(r)$ for $\alpha=1,2,3$, and $5$ in units $d^{-2}$ where $d$ is the unit used to quantify $r$.
According to (\ref{fw=eg}) a helicity measurement of a photon described by (\ref{AE-zero}) at $x^0=x^0_0$ will produce an outcome of $\pm 1$ with equal probability (of 50\%). As a result of this measurement the state of the photon collapses to the definite-helicity state whose position wave function satisfies: $\tilde f(\epsilon,\pm,\bfx)=f(\epsilon,\pm,\bfx)$ and $\tilde f(\epsilon,\mp,\bfx)=0$ (if the measured helicity value is $\pm1$.)

We end this section by noting that the quantum mechanics of a photon can be formulated in its position representation. This follows the same construction as for a (massless) scalar field that we summarize in Sec.~\ref{Sec3C}. In particular, the position wave function $f(\epsilon,\sigma,\bfx;x^0)$ for an evolving state vector $e^{-i(x^0-x^0_0)h/\hbar}\bfA$ of the photon satisfies the Schr\"odinger equation:
    \[i\hbar\partial_0 f(\epsilon,\sigma,\bfx;x^0)=\epsilon\sqrt{-\hbar^2\nabla^2}f(\epsilon,\sigma,\bfx;x^0).\]

\section{Summary and Conclusions}
\label{Sec5}

The problem of describing a first quantized photon has attracted a great deal of attention since the early 1930s. The fact that there has been no general consensus on the basic behavior of such a photon has its roots in the lack of a consistent method of constructing a corresponding Hilbert space and Hermitian operators that signify photon's observables. In the present article, we provide such a method. It consists of an explicit construction of the inner product on the space of state vectors of the photon and basic observables such as the energy, helicity, momentum, and position of the photon.

Unlike the previous studies of the subject, we do not discard the negative-energy states of the photon. To incorporate these we have introduced and provided a detailed study of the chirality (or direction-of-time) observable for the photon.

The determination of the position operator for the photon paves the way to obtain explicit form of its localized states. This in turn reveals the remarkable fact that the  electromagnetic field configuration for a localized state with definite sign of energy and helicity diverges on a plane in space. This is related to the need for an associated symmetry axis. We determine this axis and argue that depending on its choice we have different complete sets of localized state vectors with definite sign of energy and helicity. These choices are parameterized by a unit normal vector $\bfm$ to the singularity plane of these localized states and marks an implicit dependence of photon's position operator on $\bfm$. Different choices of the latter yield position operators that together with the momentum operator fulfill the canonical commutation relations. This in turn implies that they are unitary-equivalent.

Having obtained the localized states of the photon, we offer a complete solution for the notorious problem of finding a position wave function for the photon and the probability density for its  localization in space. This involves the same approach we pursue to define the position wave functions and probability density in standard non-relativistic quantum mechanics. It leads to a position representation of the quantum mechanics of a photon that is particularly useful for the calculation of the expectation values of the observables and the probability of their observation.

Among other interesting outcomes of this investigation is a systematic derivation of Hawton's position operator and a clear description of the connection between the Riemann-Silberstein and Landau-Peierls wave functions with the position wave function of the photon we introduce in this article. The latter is uniquely determined by but not identical to the electromagnetic field configuration which specifies the state of the photon. It is absolutely essential to notice that the localization of a photon in a region of space means to prepare it in a state whose position wave function vanishes outside this region. This does not imply or follow from the requirement that the corresponding electromagnetic field and its energy density should have this property.

Finally, we wish to emphasize that our approach does not rely on any ``reasonable'' or ``natural'' assumptions about the behavior of its ingredients. It is only restricted by the standard axioms of quantum mechanics and has the appealing feature of the relativistic invariance of expectation values and transition probabilities, which are the only physically measurable quantities.
\vspace{12pt}

\noindent{\bf \em Note:} After the completion of this project we came across Ref.~\cite{HB} where the authors show that by making an appropriate choice for the phase angle of the eigenvectors of the helicity operator, the localized states (in their momentum representation) acquire a definite total orbital angular momentum along their symmetry axis. \vspace{12pt}

\noindent{\bf \em Acknowledgments:} We are grateful to Farhang Loran for fruitful discussions. This work has been supported by the Turkish Academy of Sciences (T\"UBA).

\section*{Appendix~A: Calculation of vector potential and electric field for localized photons}

In this appendix we give a derivation of the formulas (\ref{LS-401}) and (\ref{LS-401-E}) for the vector potential $\bbfA^\epsilon_{\bfy,\sigma}$ and electric field $\boldsymbol{\cE}_{\bfy,\sigma}^\epsilon$ of a  photon with definite sign of energy $\epsilon$ and definite helicity $\sigma$ that is localized at a point $\bfy$ in space. This requires the evaluation of the integral on the right-hand side of Eqs.~(\ref{localized-302}) and (\ref{localized-E=}). As we mention above, we can do this only for $x^0=x^0_0$. Substituting this equation in (\ref{localized-302}) and (\ref{localized-E=}), we encounter  integrals of the form:
    \be
    \mathbf{I}_\sigma^\pm(\bfr):=\int_{\R^3}d^3\bfk\, k^{\pm 1/2}e^{i\bfk\cdot\bfr}\bfu_{\sigma}(\bfk).
    \label{Ipm=}
    \ee
We denote the components of $\mathbf{I}_\sigma^\pm(\bfr)$ by $I^\pm_j$ with $j=1,2,3$.

First we express the polarization vectors $\bfu_\sigma(\bfk)$ in the spherical coordinates in $\bfk$-space that we denote by $(k,\theta',\vartheta')$. Here $\theta'$ and $\vartheta'$ are respectively the polar and azimuthal angles, so that the components of the $\bfk$ are given by
$k_1=k\sin\theta'\cos\varphi'$, $k_2=k\sin\theta'\sin\varphi'$, and $k_3=k\cos\theta'$. Using these relations in (\ref{uk=3}), we have
    \be
    \bfu_{\sigma}(\bfk)=\frac{1-\delta_{\theta'0}\delta_{\theta'\pi}}{\sqrt2}
    \left[\begin{array}{c}
    -\cos\theta'\cos\varphi'+i\sigma\,\sin\varphi'\\
    -\cos\theta'\sin\varphi'-i\sigma\,\cos\varphi'\\
    \sin\theta'
    \end{array}\right]+
    \frac{\delta_{\theta'0}\delta_{\theta'\pi}\cos\theta'}{\sqrt2}
    \left[\begin{array}{c}
    1\\
    i\sigma\\
    0
    \end{array}\right].
    \label{pvector-spherical}
    \ee
Recall that $(r,\theta,\varphi)$ stand for the spherical coordinates of $\bfr=\bfx-\bfy$, so that the Cartesian coordinates of $\bfr$ read
    \begin{align}
    &r_1=r\sin\theta\cos\varphi, && r_2=r\sin\theta\sin\varphi, &&r_3=r\cos\theta.
    \label{sph-r=}
    \end{align}
Now, we introduce
    \begin{align}
    &\rho:=\sqrt{r_1^2+r_2^2}=r\sin\theta,
    &&a_1:=k\sin\theta'r_1,
    &&a_2:=k\sin\theta'r_2,
    \label{rho-aa}
    \end{align}
and express the integral on the right hand side of (\ref{Ipm=}) in the spherical coordinates $(k,\theta',\vartheta')$. We first compute $I^-_1$, i.e., the first component of $\mathbf{I}^-_1$. In view of (\ref{Ipm=}) and (\ref{sph-r=}) -- (\ref{rho-aa}),
    \bea
    I^-_1=
    \frac{1}{\sqrt2}\int_0^\infty dk\,k^{3/2}\int_0^\pi d\theta'\sin\theta' e^{ir_3k\cos\theta'}
    \int_0^{2\pi}d\varphi' (-\cos\theta'\cos\varphi'+i\sigma\sin\varphi')e^{i(q_1\cos\varphi'+a_2\sin\varphi')}.~~~~~~~~
    \label{Im1=}
    \eea
We evaluate the integral over $\varphi'$ to obtain $2\pi i\rho^{-1}(- r_1\cos\theta'+i\sigma r_2)J_1(k\rho\sin\theta')$, where $J_n$ stands for the Bessel J-function of order $n$ \cite{GR}. Substituting this expression in (\ref{Im1=}) and trying to evaluate the integral over $\vartheta'$ we are led to the integrals of the form
    \bea
    G_1&:=&\int_0^\pi d\theta' \sin\theta'e^{ir_3k\cos\theta'} J_1(k\rho\sin\theta')=
    -k^{-1}\partial_\rho G_3,
    \label{G1=}\\
    G_2&:=&\int_0^\pi d\theta' \sin\theta'\cos\theta'e^{ir_3k\cos\theta'} J_1(k\rho\sin\theta')=
    ik^{-2}\partial_{r_3}\partial_\rho G_3,
    \label{G2=}
    \eea
where
    \bea
    G3&:=&\int_0^\pi d\theta'e^{ir_3k\cos\theta'} J_0(k\rho\sin\theta')
    %\nn\\&=&
    =2\int_0^{\pi/2} d\theta'\cos(r_3k\cos\theta') J_0(k\rho\sin\theta')
    \nn\\&=&
    \pi\,J_0\big((r+r_3)k/2\big)J_0\big((r-r_3)k/2\big).
    \label{G3=}
    \eea
In the derivation of the latter relation we have employed formula 6.688 on page 727 of Ref.~\cite{GR}. Next, we substitute (\ref{G1=}) and (\ref{G2=}) in (\ref{Im1=}). This gives
	\be
	I^-_1= \sqrt 2\pi^2 \rho^{-1}\left(r_1 \partial_{r_3}\partial_\rho J^-+
	\sigma\, r_2 \partial_\rho J^+\right),
	\label{Im1=101}
	\ee
where $J^\pm:=\int_0^\infty dk\, k^{\pm1/2}J_0\big((r+r_3)k/2\big)J_0\big((r-r_3)k/2\big)$. With the help of formula 6.578 on page 684 of Ref.~\cite{GR}, we can evaluate the latter integral. The result is
	\begin{align}
	&J^-=\frac{\Gamma(\frac{1}{4})\, _2F_1(\frac{1}{4},\frac{1}{2};1;\sin^2\theta)}{\sqrt {2}\,\Gamma(\frac{3}{4})\, r^{1/2}},
	&& J^+=\frac{\sqrt 2\,\Gamma(\frac{3}{4})\, _2F_1(\frac{3}{4},\frac{1}{2};1;\sin^2\theta)}{\Gamma(\frac{1}{4}) \, r^{3/2}},
	\label{Jpm=}
	\end{align}
where $\Gamma$ and $ _2F_1$ stand for the Gamma and Hypergeometric functions \cite{GR}, respectively. Substituting (\ref{Jpm=}) in (\ref{Im1=101}) and using the identities $\partial_\rho=\sin\theta\,\partial_r+r^{-1}\cos\theta\,\partial_\theta$ and $\partial_{r_3}=\cos\theta\,\partial_r-r^{-1}\sin\theta\,\partial_\theta$, we evaluate the right-hand side of (\ref{Im1=101}). In view of (\ref{localized-302}), (\ref{Ipm=}),
(\ref{sph-r=}), and (\ref{rho-aa}), this gives the first component of the right-hand side of (\ref{LS-401}).

We have similarly derived the expression given by (\ref{LS-401}) for the second component of $\bbfA^\epsilon_{\bfy,\sigma}(x_0^0,\bfx)$. Our computation of the third component of $\bbfA^\epsilon_{\bfy,\sigma}(x_0^0,\bfx)$ follows the same strategy, but in addition to the above mentioned formulas of Ref.~\cite{GR} we make use of the identity:
	\be
	J_0(k\rho\sin\theta')=-(k\sin\theta')^{-2}\left(\partial^2_\rho+\rho^{-1}\partial_\rho\right)
	J_0(k\rho\sin\theta').
	\label{bessel-zero}
	\ee
This is a simple consequence of the Bessel equation: $J_0''(x)+x^{-1}J_0'(x)+J_0(x)=0$. Recall that in light of (\ref{Ipm=}), (\ref{pvector-spherical}), and the fact that $\int_0^{2\pi}d\varphi' e^{i(a_1\cos\varphi'+a_2\sin\varphi')}=2\pi J_0(k\rho\sin\theta')$, we have
	\[I^-_3=\sqrt 2\,\pi\int_0^\infty dk\,k^{3/2}\int_0^\pi d\theta'\sin^2\theta' e^{ikr_3\cos\theta'}
	J_0(k\rho\sin\theta').\]
With the help of (\ref{bessel-zero}), we can express this relation as $I^-_3=-\sqrt 2\,\pi\left(\partial_\rho^2+\rho^{-1}\partial_\rho\right)J^-$.
This together with (\ref{Jpm=}) lead us to the expression for the third component of $\bbfA^\epsilon_{\bfy,\sigma}(x_0^0,\bfx)$ as given in (\ref{LS-401}).

We have similarly used (\ref{bessel-zero}) to calculate $I^+_j$. This results in (\ref{LS-401-E}).

\section*{Appendix~B: Derivation of  $\bfu_\sigma(\bfk)$ and $\phi_\sigma^\bfm(\bfk)$}

Because $\boldsymbol{\fh}$ is a Hermitian matrix, its eigenvectors with different eigenvalues are orthogonal. We know that $\bfk$ is an eigenvector of this matrix with eigenvalue zero. This implies that the eigenvectors $\bfu_\sigma(\bfk)$ with eigenvalue $\sigma=\pm 1$ are orthogonal to $\bfk$. We can express them as linear combinations of an orthogonal pair of vectors, $\bfn_1$ and $\bfn_2$,  that lie on the plane orthogonal to $\bfk$;
	\be
	\bfu_\sigma(\bfk)=\alpha_{\sigma 1}\bfn_1+\alpha_{\sigma 2}\bfn_2,
	\label{uk-exp}
	\ee
where $\alpha_{\sigma i}$ are complex coefficients. We can express $\bfn_i$ using a real unit vector $\bfm$ that is not parallel to $\bfk$ according to $\bfn_1:=k^{-1}\bfk\times\bfm$ and $\bfn_2:=k^{-1}\bfk\times\bfn_1=k^{-2}(\bfk\cdot\bfm)\bfk-\bfm$. Substituting these relations
together with (\ref{uk-exp}) in the eigenvalue equation $\boldsymbol{\fh}\bfu_\sigma(\bfk)=\sigma\bfu_\sigma(\bfk)$, making use of the orthogonality of $\bfn_i$ and the fact that $\bfu_\sigma(\bfk)$ is a unit vector, we find $\sigma=\pm 1$ and the expression (\ref{bfuk=gen}) for $\bfu_\sigma(\bfk)$.

Next, we compute the phase factor $e^{i\phi_\sigma^\bfm(\bfk)}$ introduced in Eq.~(\ref{gauge-01}). First, we observe that according to (\ref{bfuk=gen}),
	\be
	\bfu^{\bfm}_\sigma(\bfk)=\frac{i}{\sqrt 2}(\boldsymbol{\fh}+\sigma \bone_3)\bfv^{\bfm},
	\label{app-B-001}
	\ee
where $\bfv^{\bfm}$ is a unit vector given by $\bfv^{\bfm}:=\frac{\bfk\times\bfm}{|\bfk\times\bfm|}$,
and $\bone_3$ is the $3\times 3$ identity matrix. Using (\ref{app-B-001}) to compute $\bfu^{\bbe_0}_\sigma(\bfk)$ and inserting the resulting expression together with (\ref{app-B-001}) in (\ref{gauge-01}), we deduce $(\boldsymbol{\fh}+\sigma \bone_3)[\bfv^{\bfm}-e^{i\phi_\sigma^\bfm(\bfk)}\bfv^{\bbe_0}]=0$. This identifies $\bfv^{\bfm}-e^{i\phi_\sigma^\bfm(\bfk)}\bfv^{\bbe_0}$ with an eigenvector of $\boldsymbol{\fh}$ with eigenvalue $-\sigma$. Again, because the eigenvalues of $\boldsymbol{\fh}$ are nondegenerate, we conclude that $\bfv^{\bfm}-e^{i\phi_\sigma^\bfm(\bfk)}\bfv^{\bbe_0}$ must be a constant multiple of $\bfu^{\bbe_0}_\sigma(\bfk)$. In light of (\ref{app-B-001}), this means that there is a complex number $\alpha_\sigma$ such that
	\be
	\bfv^{\bfm}-e^{i\phi_\sigma^\bfm(\bfk)}\bfv^{\bbe_0}=
	\alpha_\sigma(\boldsymbol{\fh}+\sigma \bone_3)\bfv^{\bbe_0}.
	\label{app-B-002}
	\ee
We can determine $\alpha_\sigma$ by evaluating the dot product of $\bbe_0$ with both sides of this equation and using the fact that $\bbe_0\cdot\bfv^{\bbe_0}=0$. This gives
	\be
	\alpha_\sigma=\frac{\bbe_0\cdot\bfv^{\bfm}}{\bbe_0\cdot(\boldsymbol{\fh}\bfv^{\bbe_0})}=
	\frac{ik(k_1m_2-k_2m_1)}{|\bfk\times\bfm|\,|\bfk\times\bbe_0|},
	\label{app-B-003}
	\ee
where we have employed the identity $\boldsymbol{\fh}\bfv^{\bbe_0}=ik^{-1}\bfk\times\bfv^{\bbe_0}$ which follows from (\ref{cross}). Next, we take the dot product of $\bfv^{\bbe_0}$ with both sides of (\ref{app-B-002}) and use  $\bfv^{\bbe_0}\cdot\boldsymbol{\fh}\bfv^{\bbe_0}=0$ to establish  $e^{i\phi_\sigma^\bfm(\bfk)}=\bfv^{\bbe_0}\cdot\bfv^{\bfm}+\sigma\alpha_\sigma$. This together with (\ref{app-B-003}) imply (\ref{phase=}). A highly nontrivial check on the validity of this calculation is that the right-hand side of (\ref{phase=}) is unimodular.

\section*{Appendix~C: Derivation of (\ref{fw=eg})}

First, we note that according to (\ref{AE-zero}),
	\be
	\bfB(x_0^0,\bfx)=\mathbf{0},
	\label{B-1000}
	\ee
and employ (\ref{E-zero=}) and the argument used to establish (\ref{uB=uA}) to write
$u_\sigma(\hat\bfk)^\dagger\bfE(x_0^0,\bfx)=i\sigma\hat k u_\sigma(\hat\bfk)^\dagger
\bfV(\bfx)$. Next, we substitute this relation in (\ref{f=QRS-1}) and use (\ref{B-1000}) to show that
	\be
	f(\epsilon,\sigma,\bfx)=\frac{\epsilon\,\sigma\sqrt\ell}{2(2\pi)^{3/2}}\int_{\R^3}d^3\bfk\;
	e^{i\bfk\cdot\bfx} k^{1/2}u_\sigma(\bfk)^\dagger \tilde \bfV(\bfk),
	\label{f=1001}
	\ee
where
	\be
	\tilde \bfV(\bfk):=(2\pi)^{-3/2}\int_{\R^3}d^3\bfx'\,e^{-i\bfk\cdot\bfx'}\bfV(\bfx')
	=E_0 \alpha^{-5/2}\,e^{-k^2/2\alpha}\,\bbe_0
	\label{tV=1002}
	\ee
is the Fourier transform of $\bfV(\bfx)$, and we have made use of (\ref{E-zero=}). Substituting (\ref{uk=3}) and (\ref{tV=1002}) in (\ref{f=1001}) and introducing $\cJ(\beta):=\int_0^\pi d\theta\, \sin^2\theta e^{i\beta\cos\theta}=\pi[J_0(\beta)+J_2(\beta)]/2$, we find
$f(\epsilon,\sigma,\bfx)=\epsilon\,\sigma\,\sqrt\ell E_0 [4\sqrt\pi\,\alpha^{5/2}]^{-1}
	\int_0^\infty dk\, k^{5/2} e^{-k^2/2\alpha} \cJ(k r)$.
Using Mathematica to evaluate the integral in this equation, we find (\ref{fw=eg}).


\begin{thebibliography}{99}
\bibitem{footnote1} We use the term Hermitian and self-adjoint synonymously. Note that the Hermiticity of observables is an absolutely inescapable consequence of the condition that the expectation values of observables must be real \cite{review}. A non-Hermitian operator can have a real spectrum, but there will always be states in which the expectation value of this operator is not real.
\bibitem{footnote2} Often the difficulties associated with infinite dimensional Hilbert spaces have to do with the fact that relevant operators are only defined on a proper dense subset of this space. The history of quantum mechanics teaches us that these difficulties are purely mathematical in nature. Indeed the physical state vectors that one can prepare for a measurement do not fill the whole Hilbert space.
\bibitem{footnote-2b} A linear operator $L:\cX\to\cY$ defined on $\cX$ is called an isometry if under its action the inner product of elements of $\cX$ do not change. An onto isometry is called a unitary operator.
\bibitem{footnote3} In Dirac's bra-ket notation, the symbol $\br\bfy|$ is used for $\delta_{\bfy}$.
\bibitem{footnote65} The position operators map scalar fields to scalar fields, so they do not change under Lorentz transformations. The position wave function and the probability density are determined by the inner product of pairs of scalar fields which is also Lorentz-invariant.
\bibitem{footnote6} Because $iD^{-1/2}\partial_0$ is Lorentz-invariant \cite{53}, $\bfA_c$ also corresponds to a photon field.
\bibitem{footnote7} This means that this operator maps elements of $\cH'$ to elements of $\cH'$.
\end{thebibliography}
\end{document}